\documentclass[12pt]{iopart} \usepackage{graphicx,amssymb}
\expandafter\let\csname equation*\endcsname\relax
\expandafter\let\csname endequation*\endcsname\relax
\usepackage{amsmath}
\usepackage{url}
\usepackage{hyperref}

\newcommand\far{\ensuremath{\mathcal{F}}}
\newcommand\spatialx{\ensuremath{\mathbf{x}}}
\newcommand\intparams{\ensuremath{\mathbf{\Lambda}}}
\newcommand\astprior{\ensuremath{\phi}}
\newcommand\stepfunc{\ensuremath{\Theta}}

\begin{document}

\title[PyCBC search]{The PyCBC search for gravitational waves from compact
binary coalescence}

\author{Samantha A. Usman$^{1,2}$,
        Alexander H. Nitz$^{1,3}$, 
        Ian W. Harry$^{1,4}$, \\
        Christopher M. Biwer$^1$, 
        Duncan A. Brown$^1$,
        Miriam Cabero$^3$,
        Collin D. Capano$^{3,5}$, 
        Tito Dal Canton$^3$, 
        Thomas Dent$^3$, \\
        Stephen Fairhurst$^2$, 
        Marcel S. Kehl$^{6,7}$, 
        Drew Keppel$^3$, \\
        Badri Krishnan$^3$, 
        Amber Lenon$^1$, 
        Andrew Lundgren$^3$, \\
        Alex B. Nielsen$^3$,
        Larne P. Pekowsky$^1$, 
        Harald P. Pfeiffer$^{4,6,8}$, \\
        Peter R. Saulson$^1$, 
        Matthew West$^1$,
        Joshua L. Willis$^{3,9}$.}

\address{$^1$ Department of Physics,
         Syracuse University, Syracuse, NY 13244, USA}

\address{$^2$ Cardiff University, Cardiff CF24 3AA, United Kingdom}

\address{$^3$ Max-Planck-Institut f{\"u}r Gravitationsphysik,
         Albert-Einstein-Institut, D-30167 Hannover, Germany}

\address{$^4$ Max-Planck-Institut f{\"u}r Gravitationsphysik,
         Albert-Einstein-Institut, Am M\"uhlenberg 1,
         D-14476 Golm, Germany}

\address{$^5$ Maryland Center for Fundamental
         Physics \& Joint Space-Science Institute,
         Department of Physics,
         University of Maryland, College Park, MD 20742, USA}

\address{$^6$ Canadian Institute for Theoretical Astrophysics,
         University~of~Toronto, Toronto, Ontario M5S 3H8, Canada}

\address{$^7$ Max-Planck-Institut f{\"u}r Radioastronomie, Auf dem H{\"u}gel 69,  53121 Bonn, Germany}

\address{$^8$ Canadian Institute for Advanced
  Research, 180 Dundas St.~West, Toronto, ON M5G 1Z8, Canada}

\address{$^9$ Abilene Christian University, Abilene, TX 79601, USA}

\ead{samantha.usman@ligo.org}

\begin{abstract}
We describe the PyCBC search for gravitational waves from compact-object binary 
coalescences in advanced gravitational-wave detector data.  The search was used
in the first Advanced LIGO observing run and unambiguously identified two
black hole binary mergers, GW150914 and GW151226.  At its core, the
PyCBC search performs a matched-filter search for binary merger signals
using a bank of gravitational-wave template waveforms.  We provide a complete 
description of the search pipeline including the steps used to mitigate the effects of
noise transients in the data, identify candidate events and measure their statistical 
significance.  The analysis is able to measure false-alarm rates as low as one
per million years, required for confident detection of signals.  Using data from initial
LIGO's sixth science run, we show that the new analysis reduces the background 
noise in the search, giving a $30\%$ increase in sensitive volume for binary
neutron star systems over previous searches.

\end{abstract}


\section{Introduction}
\label{s:introduction}

The detection of the binary black hole mergers GW150914 and GW151226 by the Laser
Interferometer Gravitational-wave Observatory (LIGO) has established the field
of gravitational-wave astronomy~\cite{Abbott:2016blz, 
Abbott:2016nmj, TheLIGOScientific:2016pea}.
Advanced LIGO~\cite{Abbott:2007kv,TheLIGOScientific:2016agk} will be joined by the
Virgo~\cite{Accadia:2012zzb,Acernese:2015gua} and KAGRA~\cite{Akutsu:2015hua}
detectors in the near future, forming an international network of
gravitational-wave observatories. Beyond the expected regular detections of
binary black holes~\cite{Abbott:2016nhf}, compact-object binaries containing
a neutron star and a black hole or two neutron stars are likely candidates for
detection by this network~\cite{Abadie:2010cf, Aasi:2013wya, O1:BNS_NSBH}. 
Collectively, these three types of
gravitational-wave sources are referred to as compact binary coalescences
(CBC).

The identification of gravitational-wave candidates in the detector
data is a complex process, which is performed by a \emph{search
pipeline}. The search pipeline is also responsible for determining
the significance of each identified gravitational-wave event.
This paper describes a new pipeline to search for gravitational waves 
from compact-object binaries that  uses the \emph{PyCBC} software
framework~\cite{pycbc-github} to implement an all-sky search for compact
binary coalescence. The PyCBC search pipeline described here builds upon the algorithms~\cite{Brown:2004pv,Brown:2005zs,Allen:2004gu,Allen:2005fk,Babak:2012zx}
used to search for compact-object binary coalescence with the
first-generation LIGO and Virgo
detectors~\cite{Abbott:2003pj,Abbott:2005pe,Abbott:2005kq,Abbott:2007xi,Abbott:2007ai,Abbott:2009tt,Abbott:2009qj,Abadie:2010yba,Abadie:2011nz}.
This pipeline incorporates new algorithms that improve the sensitivity of the search
and that reduce its computational cost.  We provide a complete description of
the search pipeline, emphasizing the new developments that have been made for
the advanced-detector era. To demonstrate the efficacy of PyCBC, we re-analyze
data from Initial LIGO's sixth science run and show that the new pipeline can
achieve a $\sim 30\%$ increase in sensitive volume to binary neutron stars, as
compared to the pipeline used in the original analysis of the
data~\cite{Abadie:2011nz}. Details of the results of the PyCBC search on the first aLIGO
observing run are given in \cite{TheLIGOScientific:2016qqj, O1:BNS_NSBH} with the
two observed binary black hole mergers discussed in detail in 
\cite{Abbott:2016blz, Abbott:2016nmj}.

This paper is organized as follows: Section~\ref{s:overview} provides an
overview of the search pipeline and the methods used to detect
gravitational waves from compact-object binaries in LIGO data.
Section~\ref{s:description} gives a description of the developments
implemented in this
pipeline. Section~\ref{s:ihope} compares the performance of the new pipeline
to that of the pipeline that analyzed the sixth LIGO science run and Virgo's second
and third science runs~\cite{Abadie:2011nz}. Finally,
Section~\ref{s:conclusions} summarizes our findings and suggests directions
for future improvements.
\section{Search Overview}
\label{s:overview}

The purpose of the PyCBC search is to identify candidate gravitational-wave signals 
from binary coalescences in the detector data and to provide a measure of their 
statistical significance.  Since the amplitude of the majority of gravitational-wave sources 
will be comparable to the noise background, signal processing techniques are 
required to identify the candidate events. The
gravitational waveforms for compact-object binaries can be modeled using a
combination of analytical and numerical methods \cite{Blanchet:2013haa, Taracchini:2013rva}.  
Consequently,
it is natural to use matched filtering to distinguish  signals from 
the noise background ~\cite{wainstein:1962}.  If the detector output contained
only a stationary, Gaussian noise, the matched filter signal-to-noise ratio (SNR) 
would suffice as a detection statistic since its distribution in stationary, Gaussian noise is well
known~\cite{Cutler:1992tc}. In practice, the detector data contains
non-stationary noise and non-Gaussian noise transients \cite{Aasi:2014mqd, GW150914-DETCHAR}.
Additional steps must therefore be taken to mitigate the effect of these
noise transients and to assign an accurate statistical significance to
candidate signals. Even for loud sources, the statistical
significance of candidate detections must be empirically measured since it is
not possible to shield the detectors from gravitational-wave sources and no
complete theoretical model of the detector noise exists~\cite{Creighton:1999qw}.  

\begin{figure}[tbp]
\begin{center}
\includegraphics[width=0.8\textwidth]{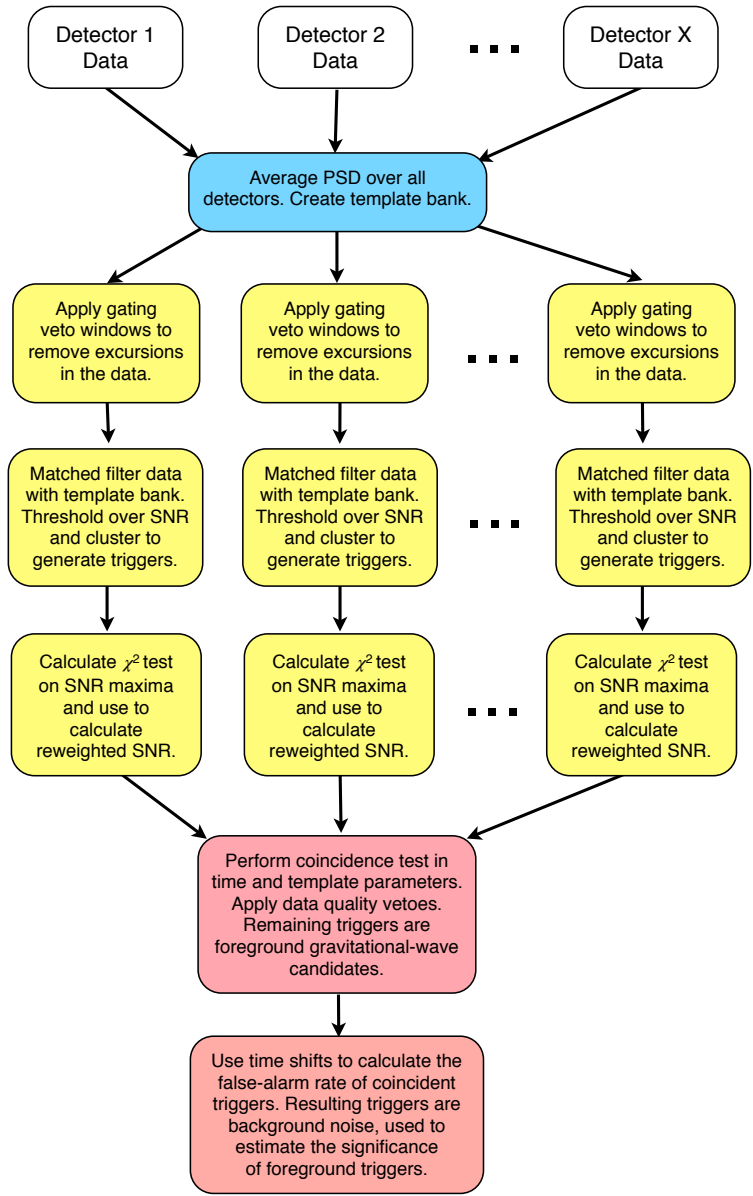} 
 \caption{A flowchart indicating the different steps of the search
pipeline. Data from the detectors are averaged to create a power spectral
density necessary to place a bank of templates that cover the search
parameter space (blue box). Times when the detector data contain loud noise
transients are removed, or \emph{vetoed}. The data from each detector is then
matched filtered and triggers are generated by thresholding and clustering the
signal-to-noise ratio time series. A chi-squared test is computed for each
trigger and the trigger's matched-filter SNR is re-weighted by the value of the 
chi-squared statistic to
better distinguish between signal and noise (yellow boxes). The pipeline
determines which triggers survive time and template coincidence, discarding
triggers that lie in times of poor data quality (red box). The triggers that
pass the coincidence and data quality tests are labelled candidate events.
Finally, multiple time shifts help generate a noise background
realization that is used to measure the significance of the candidate events
(bottom box).
\label{f:pipeline}}
\end{center}
\end{figure}

Figure~\ref{f:pipeline} shows the steps that the pipeline uses to find signals
and measure their significance. The input to the pipeline is the detector's calibrated strain
data~\cite{Siemens:2004pr,Abadie:2010px}. In addition to possible signals, the strain data contains
two classes of noise: a primarily stationary, Gaussian noise component
from fundamental processes such as thermal noise, quantum noise, and seismic
noise coupling into the detector \cite{TheLIGOScientific:2016agk} and non-Gaussian noise transients of
instrumental and environmental origin~\cite{Nuttall:2015dqa, GW150914-DETCHAR}.  To eliminate
the worst periods of detector performance, data quality
investigations~\cite{Slutsky:2010ff, Blackburn:2008ah, Aasi:2012wd, Aasi:2014mqd} 
are used to characterize detector data into three general classes: 
(i) the data is polluted with enough noise that the data should be discarded
without being searched, (ii) the data can be filtered, but candidate
events that lie in intervals of poor data quality should be discarded, or \emph{vetoed}, due to
the presence of a instrumental or environmental artifacts, or (iii) 
the data is suitable for astrophysical searches. 
Data quality
investigations are conducted independently of the search pipeline (by
looking only at detector performance), as well as by determining
the effect of instrumental artifacts on the noise background of the search
pipeline. 
 After removing data that is not suitable for
astrophysical searches, the pipeline begins its analysis, following the
remaining steps shown in Figure~\ref{f:pipeline}.

We do not \emph{a priori} know the parameters of gravitational waves in
the data, so a \emph{bank} of template waveforms is constructed that spans the
astrophysical signal
space~\cite{Sathyaprakash:1991mt, Dhurandhar:1992mw, Owen:1995tm, Owen:1998dk, Babak:2006ty, 
Harry:2009ea, Brown:2012qf, Ajith:2012mn, Keppel:2013yia, Keppel:2013uma, Privitera:2013xza, Capano:2016dsf}.
If the total mass of a compact-object binary is lower than
$M \lesssim 12\,\mathrm{M}_\odot$~\cite{Buonanno:2009zt, Brown:2012nn} and the dimensionless angular
momenta of the compact objects (their \emph{spin}) are
small $cS_{1,2}/Gm_{1,2}^2 \lesssim 0.4$~\cite{Nitz:2013mxa,Kumar:2015tha} (as
is the case for binary neutron stars, where $m_1$ and $m_2$ and $S_1$ and
$S_2$ are the component masses and spins respectively), then the detectors are
sensitive to the inspiral phase of
the waveform and this can be well modeled using the
post-Newtonian approximations (see e.g. Ref.~\cite{Blanchet:2013haa} for a
review).  For high-mass and high-spin binaries, analytic models tuned
to numerical relativity can provide accurate predictions for gravitational
waves from compact
binaries~\cite{Buonanno:1998gg,Pan:2009wj,Damour:2012ky,Taracchini:2013rva,Damour:2014sva}.
The template bank is constructed to cover the space of circular binaries with aligned spins.  It
is generated so that the loss in matched-filter SNR due to the
discrete nature of the bank is no more than $3\%$. The exact placement
of the templates depends on the detector's noise power spectral density (PSD) $S_n(f)$.  
The PyCBC search pipeline places templates using a single noise PSD, averaged
over all of the time and all detectors in the search network.  The data from
each detector in the network is matched filtered against this single template bank.
We describe the process of constructing this template bank in Section~\ref{s:bank}.
 
Since the waveform of the target signals are well modeled, the pipeline uses
matched filtering to search for these signals in detector noise. In the PyCBC
pipeline, each detector's data is filtered independently.  The search templates 
are restricted to spin aligned binaries, and only the dominant gravitational wave harmonic
\cite{Capano:2013raa}.  Consequently, the sky location and orientation of the binary
affect only the overall amplitude and phase of the waveform.  These are maximized
over when constructing the matched filter by projecting the data signal
against two orthogonal phases of the template $h(t)$, given by
$h_\mathrm{cos}$ and $h_\mathrm{sin}$~\cite{Allen:2005fk}.  The matched filter
then consists of a weighted inner product in the frequency domain used to
construct the signal-to-noise ratio (SNR), $\rho(t)$, as:
\begin{equation}
\rho^2(t) = \frac{(s|h_\mathrm{cos})^2}{(h_\mathrm{cos}|h_\mathrm{cos})} + \frac{(s|h_\mathrm{sin})^2}{(h_\mathrm{sin}|h_\mathrm{sin})} = \frac{(s|h_\mathrm{cos})^2 + (s|h_\mathrm{sin})^2}{(h_\mathrm{cos}|h_\mathrm{cos})},
\label{eq:snr}
\end{equation}
where the inner product is given by
\begin{equation}
(s|h)(t) = 4\mathrm{Re}\int_{f_\mathrm{low}}^{f_\mathrm{high}} \frac{\tilde{s}(f)\tilde{h}^*(f)}{S_n (f)}e^{2\pi i f t}\, \mathrm{d}f.
\label{eq:ip}
\end{equation}
Here $\tilde{s}(f)$ denotes the Fourier-transformed detector data, defined by
\begin{equation}
\tilde{s}(f) = \int_{-\infty}^{+\infty}s(t)e^{-2\pi itf}dt,
\end{equation}
and $\tilde{h}(f)$ denotes the Fourier-transformed template waveform. $S_n(f)$
is the one-sided PSD of the detector noise defined by
\begin{equation}
\langle\tilde{s}(f) \tilde{s}(f')\rangle = \frac{1}{2} S_n(f) \delta(f-f'),
\end{equation}
where angle brackets denote averaging over noise realizations and $\delta$ is
the Dirac delta function. The frequency limits
$f_\mathrm{low}$ and $f_\mathrm{high}$ are determined by the bandwidth of the
detector's data, and the two phases of the template are
related by $\tilde{h}_\mathrm{sin}(f) = i \tilde{h}_\mathrm{cos}(f)$.

Despite extensive data quality investigations, noise transients of unknown
origin still remain after data quality vetoes are applied to the search.
To mitigate the effect of these noise transients, the
pipeline identifies excursions in the input strain data $s(t)$ and then
applies a window to the detector data, zeroing out the data around the time of
a noise transient before filtering. This procedure, called \emph{gating}, is
described in Section~\ref{s:gating}. Having removed these noise transients,
the pipeline computes the matched-filter SNR $\rho(t)$ for each template in
each detector.  The search identifies the times when the matched-filter SNR
exceeds a predetermined threshold for a given template in a detector's data.
The pipeline applies a clustering algorithm, which takes the largest value
within a predefined window of $\rho(t)$ and identifies maxima in the SNR time
series. This process yields a list of times when a signal may be present, which are called
\emph{triggers}.  The matched filtering, thresholding and clustering
algorithms are described in Section~\ref{s:filtering}. 

Triggers generated by matched filtering the data against the template bank are
subject to a chi-squared test that determines if the time-frequency
distribution of power in the data is consistent with the expected power in the
matching template waveform~\cite{Allen:2004gu}.  To construct this test, the
template is split into $p$ frequency bins. These bins are constructed so that
each contributes an equal amount of power to the total matched-filter SNR. The
matched-filter SNR, $\rho_i$, is constructed for each of these $p$ bins. For a
real signal, $\rho_i$ should contain $1/p$ of the total power. The $\chi^2$
statistic compares the expected to the measured power in each bin according to
\begin{equation}
\chi^2 = p \displaystyle\sum_{i=1}^{p}\left[
\left( \frac{\rho_\mathrm{cos}^2}{p}-\rho_{\mathrm{cos},i}^2\right)^2 + 
\left( \frac{\rho_\mathrm{sin}^{2}}{p}-\rho_{\mathrm{sin},i}^2\right)^2 \right] \, ,
\label{eq:chisqr}
\end{equation}
where $\rho_\mathrm{cos}^2$ and $\rho_\mathrm{sin}^2$ are the SNRs of the two
orthogonal phases of the matched filter.  Lower-mass binary systems, such as
binary neutron stars, lose energy to gravitational waves more slowly than
higher-mass systems. Consequently, the waveforms of lower mass systems are
longer, having more gravitational-wave cycles in the sensitive band of the
detector.  
The PyCBC pipeline allows
the number of bins to be specified as a function of the intrinsic parameters
of the template.  This allows the search to use more bins in the chi-squared
test for longer templates, making the test more effective.  In previous analyses,
the chi-squared test was the most computationally costly part of the pipeline 
\cite{Babak:2012zx}.  The PyCBC pipeline uses a more efficient algorithm
for computing the chi-squared statistic, which vastly reduces the computational
cost.  Details of the chi-squared test are given in Section~\ref{s:chisq}.

For a trigger of given matched-filter SNR, larger values of $\chi^2$ indicate a higher likelihood 
of a noise transient origin as opposed to a signal. For signals, the reduced chi-squared,
$\chi_r^2 = \chi^2/(2p-2)$, should be near unity.  To down-weight triggers 
caused by noise transients, the matched-filter SNR is 
re-weighted~\cite{Abadie:2011nz,Babak:2012zx} according to
\begin{equation}
\hat{\rho} = \left\{\begin{array}{lr}
\rho\left/\left[(1+(\chi^2_r)^3)/2\right]^\frac{1}{6}\right., & \text{if } \chi_r^2 > 1, \\
\rho, & \text{if } \chi_r^2 \le 1.
\label{eq:reweighted}
\end{array}\right.
\end{equation}
Having computed the re-weighted SNR for each trigger, the pipeline discards
all triggers that lie below a pre-determined re-weighted SNR threshold.
 
The search requires that signals are observed with consistent parameters in
the detector network.  First, any triggers that occur during times
of instrumental or environmental artifacts, as determined by the input data
quality metadata, are \emph{vetoed}. 
To be
considered a candidate event, triggers must be observed with a time of arrival
difference less than or equal to the gravitational-wave travel time between
detectors, with an additional window to account for uncertainty in the
measurement of the time of arrival. The triggers must also be observed with
the same template in both detectors. Triggers that survive the time and
parameter coincidence test are referred to as \emph{candidate events}.  Details
of the coincidence tests are presented in Section~\ref{s:coinc}. 

The quadrature sum of the
reweighted SNR $\hat{\rho}$ in each detector is the pipeline's \emph{detection statistic}
which ranks the likelihood that a trigger is due to a gravitational-wave signal. 
To assign a statistical significance to detection candidates, the pipeline
measures the false-alarm rate of the search as a function of the
detection-statistic value $\hat{\rho}_c$. Since it is not possible to isolate
the detectors from gravitational waves, it is impossible to directly measure
the detector noise in the absence of signals. This, together with the
non-stationary and non-Gaussian nature of the noise, means that the
false-alarm rate of the search must be empirically measured.  This is done by
applying a time shift to the triggers from one detector relative to
another.  The minimum time-shift offset is chosen to
be larger than the time-coincidence window used to determine if signals are
observed with consistent parameters in the network. Events in the time-shifted
analysis therefore cannot be due to the coincidence of the pair of triggers
produced by a real gravitational-wave signal.
Many time shifts create a large background data set which are used to 
approximate the background noise and estimate the search's false-alarm rate.

Since different templates in the bank can respond to detector noise in 
different ways, the search background is not uniform across the template bank.  
To maintain the sensitivity of the search to signals over a wide range of 
masses under this non-uniform background distribution, the search sorts 
both candidate events and background events into different classes.  The 
false-alarm rate of the search in each class is used to assign a p-value to the 
candidate events; a given candidate event is compared to the background events 
from the same class.
To account for having searched multiple classes whose response is not
completely independent~\cite{Lyons:1900zz}, the significance of candidate 
events is decreased by applying a trials factor equal to the number of bins, 
$n_\mathrm{bins}$ to obtain a final p-value which describes the statistical 
significance of a candidate event. This procedure is described in 
Section~\ref{s:far}.

\section{Search Description}
\label{s:description}

In this section we describe the methods and algorithms used in the PyCBC
search pipeline introduced in Section~\ref{s:overview}.  
In particular, we focus on the parts of the PyCBC search that
improve upon the ihope pipeline described in Ref.~\cite{Babak:2012zx}.

\subsection{Template Bank Placement}
\label{s:bank}

Methods for creating template banks of non-spinning and aligned-spin
waveform filters for a given parameter space have been extensively explored in the
literature~\cite{Sathyaprakash:1991mt,Dhurandhar:1992mw,Owen:1995tm,Owen:1998dk,Babak:2006ty,Harry:2009ea,Brown:2012qf,Ajith:2012mn,Keppel:2013yia,Keppel:2013uma,Privitera:2013xza,Capano:2016dsf,Harry:2013tca}.
Here we use the method presented in~\cite{Capano:2016dsf}.  The density of
templates across the parameter space depends upon the noise power spectrum of
the detectors.

The PyCBC search generates a template bank covering the four dimensional space of mass
and aligned spin of the two components. The pipeline  
uses a single template bank for all detectors and over the entire duration of the search. Using a common
template bank for all templates allows us to require coincident events to be observed in the same
template in different detectors, as discussed in Section~\ref{s:coinc}, which improves the sensitivity of
the pipeline.

To generate an appropriate bank, we must compute a noise PSD
estimate averaged both over time and over detectors.  
We have explored several different methods to create a single noise PSD
that is valid for the duration of the analysis period~\cite{Kehl:2014} and find that
the harmonic mean provides the best estimate for placing
the template bank. Specifically, we measure the noise PSD
every 2048~seconds over the observation period, independently in each detector, using
the median method of Ref.~\cite{Allen:2005fk}. We then obtain $N_s$ power spectra $S_n$ for each detector
in the network. We first construct the harmonic mean PSD for a single
detector, defined by averaging each of the $f_k$ frequency bins independently according to
\begin{equation}
S_n^\mathrm{harmonic}(f_k) = N_s \left/ \sum_{i=1}^{N_s}\frac{1}{S_n^i(f_k)} \right.\, .
\end{equation}
We then use the same method to compute the harmonic mean of the resulting PSDs
 from each detector in the network.

The PSD estimate, and hence the bank itself, only needs to be re-generated 
when there
are significant changes in the detector's noise PSD. This typically
only happens when significant physical changes occur in the detectors. Using a single
PSD estimate for an extended period of time allows the efficient use
of template banks that include compact-object spin, as demonstrated in
Ref.~\cite{TheLIGOScientific:2016qqj}. In Section 4, we compare
the sensitivity and the computational cost of using a single template bank constructed
with an averaged PSD and that of using regenerated template banks
with shorter PSD samples.

\subsection{Removal of Non-Gaussian Noise Prior to Filtering}
\label{s:gating}

Transient noise in the detectors' data streams can produce high SNR triggers, even
when the transients do not resemble the templates.  Data quality investigations and
vetoes~\cite{Aasi:2012wd, Aasi:2014mqd} remove many, but by no means all, of these
loud transients which can affect the astrophysical
sensitivity of the search. Although short-duration, loud transients (or \emph{glitches}) that
survive data-quality investigations are suppressed by the chi-squared test,
they can reduce search sensitivity through two mechanisms: \emph{dead time}
due to the clustering algorithm and \emph{ringing} of the matched filter; both
of these mechanisms are related to the impulse response of the matched filter.
In this section, we explain the origin of these features and describe the
methods used by the pipeline to reduce their effect on the search's sensitivity.

\begin{figure}[t!bp]  
\begin{center}
\includegraphics[width=0.8\linewidth]{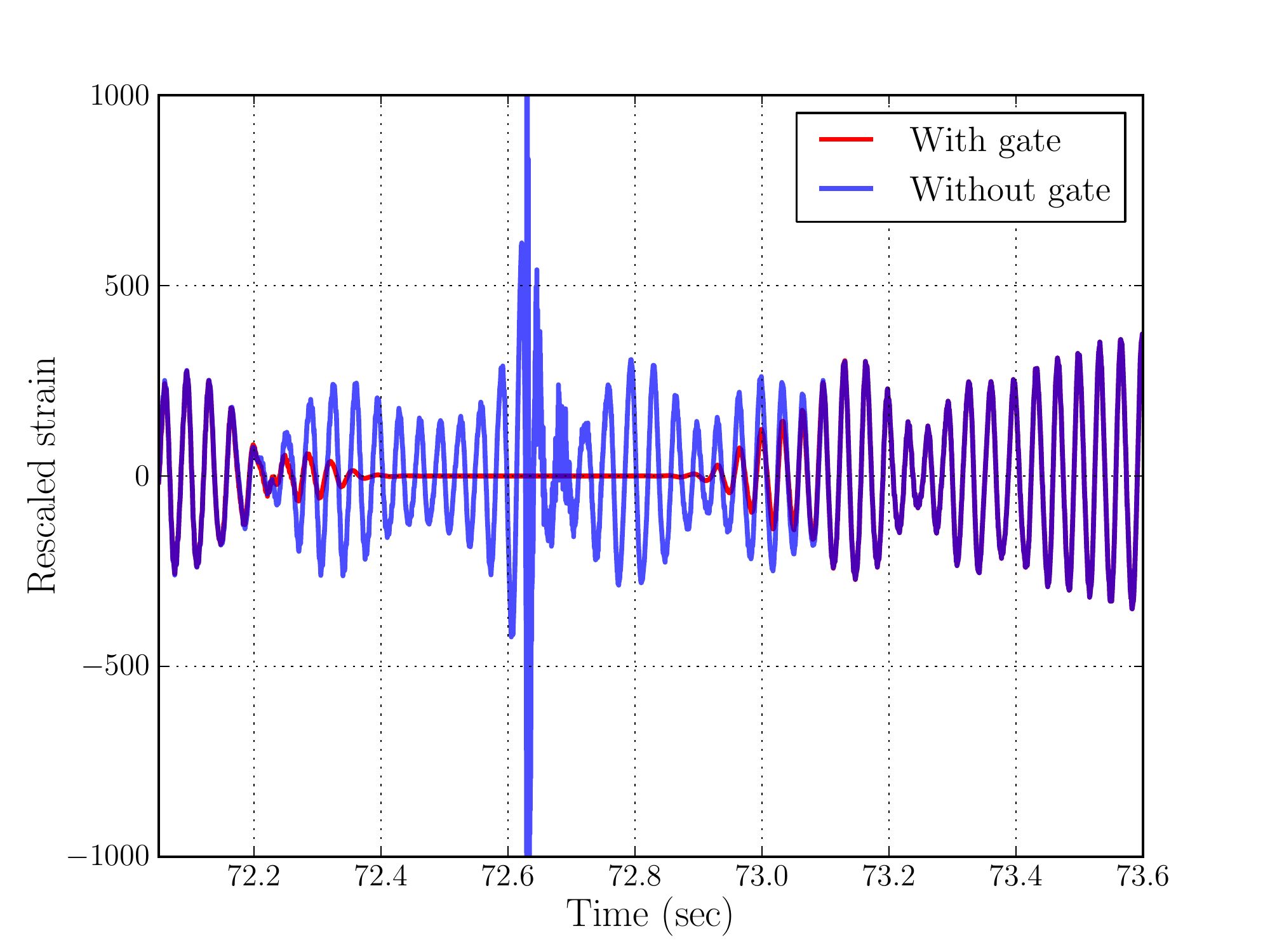}
\caption{The effect of gating on a loud noise transient. Both lines show the detector strain, which has been rescaled by a factor of
$10^{21}$ prior to filtering. On this scale, the transient has a peak
magnitude over 5,000. The blue line shows the data before applying
the Tukey window, the red line shows the data after.}
\label{gated-strain}
\end{center}
\end{figure}

The impulse response of the matched filter is obtained by considering a
delta-function glitch in the input data $s(t) = \delta(t - t_g)$, where $t_g$
is the time of the transient. Although not all types of glitches are of this
nature~\cite{Canton:2013joa}, many loud noise transients are well
approximated as $s(t) = n(t) + \delta(t - t_g)$. For example,
Figure~\ref{gated-strain} shows a typical loud transient glitch from LIGO's
sixth science run.
As loud noise transients occur frequently in detector data, most glitches are
not individually investigated.  Although the cause of this glitch is unknown,
it likely comes from an instability in the detector's control systems.
For such a glitch, the
matched-filter SNR given by Eq.~(\ref{eq:snr}) will be dominated by
\begin{equation}
\rho^2(t) \approx I_\mathrm{cos}^2(t - t_g) + I_\mathrm{sin}^2(t - t_g),
\end{equation}
where
\begin{eqnarray}
I_{\mathrm{cos},\mathrm{sin}}(t - t_g) = 4\mathrm{Re} \int_{f_{\mathrm{low}}}^{f_{\mathrm{high}}}
 \frac{\tilde{h}_{\mathrm{cos},\mathrm{sin}}(f)}{S_n(f)} e^{2\pi i f (t_g - t)} \, df
\end{eqnarray}
is the impulse response of each of the two phases of the matched filter.  This is
equal to the convolution of the template, $h_{\mathrm{cos},\mathrm{sin}}(t)$ with
the inverse Fourier transform of inverse power spectral density, $1 / S_{n}(f)$.  
To ensure that the impulse response of the filter is of finite duration, the inverse PSD 
is truncated to a duration of 16~seconds in the time domain before
filtering~\cite{Allen:2005fk}.  The inverse PSD truncation serves to smear out very sharp
features in the PSD: using a 16-second window provides sufficient resolution of these features,
while minimizing the length of the impulse response.  The length of the template can be on
the order of minutes, leading to very long impulse response times for the
matched filter, although the template amplitude is quite sharply peaked near merger.

\begin{figure}[!tbp]  
\begin{center}
\includegraphics[width=0.8\linewidth]{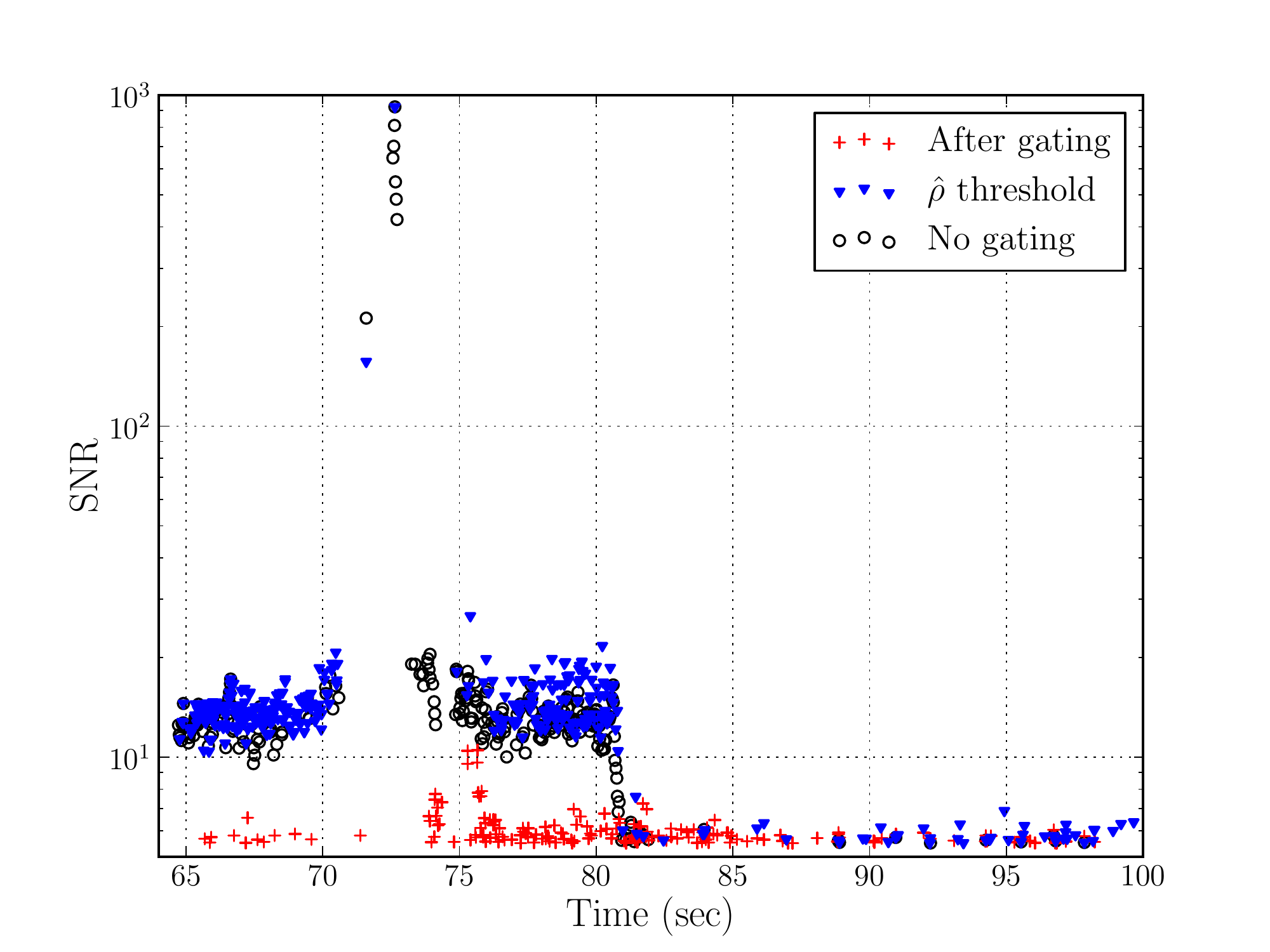}
\caption{Response of the pipeline to a loud glitch. The matched-filter SNR of
the triggers generated is shown as a function of time. Black circles
show the triggers generated immediately after filtering without gating applied
to the input data. Blue triangles show the triggers that remain after the
re-weighted SNR is computed for each trigger and a threshold of 
$\hat{\rho} > 5.5$ is applied. Although the significance of these triggers is
suppressed by the re-weighted SNR, many triggers still remain which can
increase the noise background of the search. The red crosses show the
triggers produced by the search after gating is applied and the SNR is
reweighted. The large majority of triggers caused by the glitch in the $\pm
8$~seconds around the transient are absent.}
\label{triggers-after-glitch}
\end{center}
\end{figure}

Figure~\ref{triggers-after-glitch} shows the result of filtering the glitch in
Figure~\ref{gated-strain} through a template bank and generating triggers from
the matched-filter SNR time series. The black circles show the triggers
generated by the ringing of the filter due to the glitch. The largest SNR values
occur close to the time of the glitch and are due to the impulse response of
the template $h$. The shoulders on either side of this are due to the impulse
response of the inverse power spectrum $1/S_n(f)$.   This leads to the two effects 
described above: an excess of
triggers around the time of the glitch which can increase the noise background
of the search and a window of time containing multiple noise triggers that can
make it difficult to distinguish signal from noise. 
In Figure~\ref{triggers-after-glitch}, we also plot the triggers that have
a re-weighted SNR above the threshold $\hat{\rho} > 5.5$.  Although the use of the
chi-squared veto to construct the reweighted SNR suppresses the significance of
these triggers, it does not completely remove them from the analysis and 
an increased trigger rate around the time of the glitch remains.

The PyCBC pipeline identifies non-stationary transients and removes them
from the data, a process called \emph{gating}. The data around short-duration noise transients
is set to zero prior to matched filtering. To zero the data, the input data $s(t)$ is multiplied by a
window function centered on the time of the peak. A Tukey window is applied to
smoothly roll the data to zero and prevent discontinuities in the input data.
The Tukey window used has a shape parameter of $\alpha =
0.5$, where $\alpha$ is the fraction of the window inside the cosine tapered
region.
The effect of this gating on the strain data is shown in
Figure~\ref{gated-strain} where the input data is zeroed for $1$~second around the
time of the glitch. Figure~\ref{triggers-after-glitch} shows the output of the search
when run on the gated data.  
In addition to removing the loud triggers with SNR $\sim 1000$ at the
time of the glitch, gating removed the additional triggers with SNR $\sim 10$
generated by the impulse response of the filter before and after the glitch. This
improves search sensitivity by reducing the amount of data corrupted by the
glitch to only the windowed-out data, and reduces the overall noise background
by removing noise triggers that could possibly form coincident events.
Gating is typically only applied to short-duration transients length of order
$1$~s. Longer duration transients that are identified by data quality
investigations are removed prior to analysis by the pipeline.

This process requires the identification of the time of noise transients that
will be removed by gating. These times can be determined by either data 
quality investigations or by a separate search 
for excess power in the input strain data (typically with a high SNR 
threshold)~\cite{Chatterji:2004qg,Robinet:2015om}; this method
was used in Ref.~\cite{Abbott:2016blz}. Alternatively, short-duration glitches
that are not flagged by data quality investigations can be identified by the pipeline.
To do this, the pipeline measures the power spectral density of the input time
series. The data is Fourier transformed, whitened in the frequency domain, and
then inverse Fourier transformed to create a whitened time-series. The
magnitude of the whitened strain data is computed and times exceeding a
pre-determined threshold are identified. Peaks that lie above the
threshold are identified by the time-clustering algorithm of
Ref.~\cite{Allen:2005fk}. The data around these times is then gated
using the procedure described above.

\subsection{Matched Filtering}
\label{s:filtering}

A core task of the search pipeline is to correlate the detector data against a
bank of template waveforms to construct the matched filter signal-to-noise
ratio. The matched filtering used in our search pipeline is based on the
FindChirp algorithm developed for use in the Initial LIGO/Virgo searches for
gravitational waves from compact binaries~\cite{Allen:2005fk}. In this section
we describe two improvements to the FindChirp algorithm: changes to the
noise power spectral density estimation and a new thresholding
and clustering algorithm that identifies maxima in the signal-to-noise ratio
time series.

The matched filter in Eq.~(\ref{eq:ip}) is typically written in terms of
continuous quantities, e.g. $s(t), h(t)$, and $S_n(f)$. In practice,
the pipeline works with discretely sampled quantities, e.g. $s_j \equiv
s(t_j)$, where $s_j$ represents the value of $s(t)$ at a particular time
$t_j$. Similarly, $S_n(f_k)$ represents the value of the noise power spectral
density at the discrete frequency $f_k$. The input strain data is a discretely
sampled quantity with a fixed sampling interval, typically $\Delta t =
1/4096$\,s. Fourier transforms are computed using the Fast Fourier Transform
(FFT) algorithm in blocks of length $T_B = 256$~seconds. 
For this block length, the number of discretely sampled data points in
the input time-series data $s_j$ is $N = T_B / \Delta t = 256 \times 4096 =
2^{20}$. The discrete Fourier transform of $s_j$ is given by
\begin{equation}
\tilde{s}_k = \sum_{j=0}^{N-1} s_j e^{-2 \pi ijk/N},
\end{equation}
where $k = f_k / (N \Delta t)$. This quantity has a frequency resolution given
by $\Delta f = 1 / (N\Delta t)$. 

To construct the integrand of the matched filter, the discrete quantities
$\tilde{h}_k$, $\tilde{s}_k$, and $S_n(f_k)$ must all have the same length and
frequency resolution.  The most straightforward way to do this is to
couple the computation of $\tilde{s}_k$ and $S_n(f_k)$, by using the same
length of data to compute both.  An average PSD is then computed by
averaging (typically) 15 overlapping blocks of 
data~\cite{Brown:2004vh,Allen:2005fk,Babak:2012zx},
leading to a large variance in the estimated PSD. 

The PyCBC pipeline decouples the computation of $S_n(f_k)$ and $\tilde{s}_k$,
allowing many more averages to be used to compute $S_n(f_k)$.  As discussed
in Section \ref{s:gating}, the inverse PSD is truncated to 16~seconds to reduce
the length of the filter's impulse response: thus, it is natural to estimate the 
PSD using $16$-second time-domain blocks.  An input filter length of $2048$~seconds allows 
for $127$ such blocks, leading to significantly less variance in the PSD estimate. 
The resulting PSD is then linearly interpolated to match the resolution of the 
data.  Finally, the interpolated $S_n(f_k)$ is
inverted, inverse Fourier transformed, truncated to a fixed length of
$16$~seconds
in the time domain~\cite{Allen:2005fk}, and then Fourier transformed to the 
frequency domain for use in the matched filter. Implementation of the matched
filtering algoritm in PyCBC can be performed using either single-threaded or
parallel FFT engines, such as FFTW~\cite{FFTW05} or the Intel Math Kernel
Library. The choice of single- or multi-threaded filtering is made at runtime,
so that the fastest implementation (one multi-threaded process, or several
single-threaded processes) can be made depending on the architecture used.

Once the matched-filter SNR time series $\rho^2_j$ has been computed, the
final step of the filtering is to generate \emph{triggers}. These are maxima
where the SNR time series exceeds a chosen threshold value.
For either signals or noise transients, many sample points in the
SNR time series can exceed the SNR
threshold. Since a real signal will have a single, narrow peak in the SNR time
series, the pipeline applies a time-clustering algorithm to keep local maxima of the 
SNR time series that exceed the threshold. 

The PyCBC pipeline divides the 256-second SNR time series 
into equal 1-second windows and then identifies the maximum of the
time series within each window. As the maximization within each window is
independent from other windows, the clustering for each window can be parallelized
for efficiency on multi-core processing units.
If the desired clustering window is greater than 1 second, then the initial
list of maxima---potentially one for each window within the segment---is
further clustered: a candidate
trigger in a window is kept only if it has a higher SNR than
both the window before and after it, ensuring that it is a local maximum in
the SNR time series. This clustering algorithm is more computationally efficient
than the running maximization over template length described in Ref.~\cite{Allen:2005fk}.
Furthermore, it reduces the clustering window, thereby increasing the search sensitivity.

\subsection{Signal Consistency Tests}
\label{s:chisq}

The chi-squared signal-consistency test introduced in Ref.~\cite{Allen:1999yt}
and developed in Ref.~\cite{Allen:2004gu} is a powerful way to distinguish
between signals and noise in searches for gravitational waves from
compact-object binaries.  The PyCBC search allows the number of bins $p$ in
Eq.~(\ref{eq:chisqr}) to be varied as a function of the parameters of the
template~\cite{TheLIGOScientific:2016qqj}.  This results in an improvement in
search sensitivity, as demonstrated in Section~\ref{s:ihope}.

The chi-squared statistic can be computationally intensive to calculate, however. 
For every FFT operation required to compute the SNR time series, the
chi-squared statistic requires an additional $p$ FFT operations, one for each frequency bin of
Eq.~(\ref{eq:chisqr}).  The PyCBC pipeline instead calculates the chi-squared test only
at the time samples corresponding to clustered triggers in the matched-filter
SNR time series. To do this, the pipeline uses an optimized integral over
frequency, rather than computing the FFT. If data quality is poor and the
number of triggers in a FFT block is large, the FFT method becomes more
efficient. We quantify this cross-over point below.

To compare the computational cost of the two methods, we consider the
calculation of the $p$ matched filters for the na{\"i}ve chi-squared test
implementation.  Computation of the $\rho^2_i$ for each of the $p$ bins
requires an inverse complex FFT. For a data set containing $N$ sample points
the number of operations is $p \times 5 N \log(N).$
\footnote{Since the FFT dominates
the operations count, we have neglected lower-order terms that do not
significantly contribute to the computational cost.}  
If we instead calculate the chi-squared value only at the peaks, we must 
evaluate 
\begin{equation}
\frac{\chi^2 + \rho^2}{p}[j] = \sum_{i=1}^{p}\rho_i^2 [j],
\label{eq:chisq-comp}
\end{equation}
at the set of points $[j]$ identified as triggers.  We can re-write Eq.~(\ref{eq:chisq-comp}) as
\begin{equation}
\frac{\chi^2 + \rho^2}{p}[j] = \sum_{i=1}^{p}
\left(\sum_{k=k^\mathrm{min}_i}^{k^\mathrm{max}_i}\tilde{q}_k e^{2\pi i j k/N}\right)^2\, ,
\label{eq:twiddle}
\end{equation}
where $k_i^\mathrm{min,max}$ denote the frequency boundaries of the $p$ bins
and are given by $k_i^\mathrm{min,max} = f_i^\mathrm{min,max} / \Delta f$, and 
$\tilde{q}_{k}$ is the kernel of the matched filter, defined as~\cite{Brown:2004vh}
\begin{equation}
\tilde{q}_{k} = \frac{\tilde{s}(f) \tilde{h}(f) }{S_n(f)} \, .
\end{equation}
We can further reduce the computational cost of the chi-squared by noting that the 
exponential term requires the explicit calculation of $e^{2\pi i j k/N}$ at 
$k_{\mathrm{max}} = N/2$ points. This can be
reduced to a single computation of the exponential term by pre-calculating 
$e^{2\pi i j/N}$ once and then iteratively multiplying by this constant to
obtain the next term needed in the sum. 
To do this, we write Eq.~(\ref{eq:twiddle}) in the following form:
\begin{equation}
\frac{\chi^2 + \rho^2}{p}[j] = \sum_{i=1}^{p}
\left( \sum_{k=k^{min}_i}^{k^\mathrm{max}_i}\tilde{q}_k e^{2\pi i j/N} (e^{2\pi i j/N})^{k-1} \right)^2 \, .
\label{eq:fasttwiddle}
\end{equation}
This reduces the computational cost of each term in the sum to two complex multiplications: one 
to multiply by the pre-computed constant $e^{2\pi i j/N}$ and one for the multiplication 
by $\tilde{q}$. Computing the right hand side of Eq.~(\ref{eq:fasttwiddle}) then
requires the addition of two complex numbers for each term in the sum. The
total computational cost to compute the chi-squared test for $N_P$ points is
then $14 k_\mathrm{max} \times N_P = 7N \times N_P$. 

For small numbers of matched-filter SNR threshold crossings, this new
algorithm can be significantly less costly than calculating the chi-squared
statistic using the FFT method. However, if the number of threshold crossings
$N_P$ is large, then the FFT method will be more efficient due to the $\log N$
term. The crossover point can be estimated for $p$ chi-squared bins as
\begin{equation}
N_p = \frac{p \times 5N \log(N)}{14 k_\mathrm{max}} = \frac{5}{7} \, p \log(N), 
\label{eq:chisq-points}
\end{equation}
although this equation is approximate because the computational cost of an FFT
is highly influenced by its memory access pattern. For a typical LIGO search
where $N = 2^{20}$, the new algorithm is more efficient when the number of
points at which the $\chi^2$ statistic must be evaluated is $N_p \lesssim
100.$ For real LIGO data, the number of times that the $\chi^2$ statistic must
be evaluated is found to be less than this threshold on average,
and so this method significantly reduces the computational cost of the pipeline. However, there are
still periods of time where the data quality is poor and the FFT method is
more efficient. Consequently, the pipeline described here computes $N_P$ and
uses either the single-trigger or FFT method, depending on the threshold
determined by Eq.~(\ref{eq:chisq-points}). For a typical analysis configuration that only 
identifies triggers once every second, this threshold is never exceeded, and the single-trigger 
method is faster even where the data quality is poor.

\subsection{Coincidence Test}
\label{s:coinc}

To reduce the rate of false signals, we require that a candidate event is seen
with consistent parameters in all of the detectors in the network. The pipeline
enforces this requirement by performing a \emph{coincidence test} on triggers
produced by the matched filter. This test is performed after triggers occuring
during times affected by known instrumental or environmental artifacts have been 
discarded. The PyCBC pipeline requires consistency of arrival time and template
parameters (masses and spins)
between different detectors.  Here, we only consider a two-detector network, but
the extension of this coincidence test to more than two detectors is straightforward.

For the two-detector LHO-LLO network, signals must be seen in both detectors 
within a time difference of $15$\,ms: $10$\,ms maximum travel time between
detectors and $5$\,ms padding to account for timing errors.  
Since the same waveform should be observed in all detectors, the pipeline requires consistency 
of the parameters of the best-fit template between detectors. 
Previous searches used a metric-based coincidence test~\cite{Robinson:2008un}
that checked the consistency of the template parameters and arrival times of
triggers between detectors, but did not require template parameters to be the same.  
The PyCBC pipeline requires an \emph{exact-match} coincidence. The same
template bank is used to filter data and produce triggers from each detector. In addition to
enforcing coinsistent arrival times, the intrinsic parameters (masses and spins) of
triggers in each of the detectors must be exactly the same.  The
exact-match coincidence test is useful in cases where there is no simple
metric to compare gravitational waveforms, such as template waveforms for
binaries with spinning neutron stars or black holes~\cite{Canton:2014ena} or
for high-mass waveforms where a stochastic template placement algorithm is
used~\cite{Capano:2016dsf,TheLIGOScientific:2016qqj}.  As demonstrated in
Section \ref{s:ihope}, the use of exact match coincidence leads to a
measurable improvement in the search sensitivity, even when a metric-based
coincidence test is available.

Triggers that survive the coincidence test are considered coincident events. These 
candidates are then ranked by the quadrature sum of the reweighted SNR in each 
detector; for a two-detector network, we have
\begin{equation}
\hat{\rho}_c = \sqrt{\hat{\rho}_\mathrm{1}^2 + \hat{\rho}_\mathrm{2}^2}.
\end{equation}

The choice of coincidence test has implications for the number of 
single-detector triggers that must be stored.  
In previous searches, triggers were clustered over the entire template 
bank~\cite{Robinson:2008un,Babak:2012zx} prior to generating coincident
events.  However, applying such clustering in conjunction 
with exact-match coincidence leads to unacceptable reductions in search 
sensitivity, as random noise causes the highest-SNR template to vary 
between detectors for quiet simulated signals.  Thus, we keep all triggers
from the bank before testing for coincidence. 

\subsection{Candidate Event Significance}
\label{s:far}

The final step of the pipeline is to measure the false-alarm rate of the
search as a function of the detection statistic, $\hat{\rho}_c$, and use this
to assign a statistical significance to candidate events. The rate and
distribution of false alarms in the search, i.e. coincident triggers due to 
noise, depends on the pipeline's response 
to unpredictable, non-Gaussian and non-stationary detector noise and 
must be measured empirically. The PyCBC pipeline measures the false-alarm rate 
using time shifts.  In this section, we describe the method in detail, again
restricting attention to the two detector case.  Generalisations to additional
detectors are relatively straightforward.

Triggers from one detector are shifted in time with respect to
triggers from the second detector, then the coincidence test is re-computed 
to create a background data set that does not contain coincident
gravitational-wave signals. Repeating this
procedure many times on a sufficiently large duration of input data produces a 
large sample of false coincidences that are used to compute the 
false-alarm rate of the search as a function of the detection statistic. 
Under the assumptions that the times of transient
noise artifacts present in the data streams are not correlated between
different detectors and that gravitational-wave signals are sparse in the
data set, this is a sufficiently accurate approximation of the background to 
support detection claims. 

Our significance calculation requires that candidate events are statistically 
independent; however, both noise transients and signals can generate many triggers 
across the template bank, potentially yielding several correlated coincident
events within a short time ($\lesssim 1$~second). To generate independent
candidates, the pipeline performs a final stage of clustering: if more than one
coincident event occurs within a time window of fixed duration, 
typically 10~seconds, only the event with the highest detection statistic value 
An identical clustering operation is also performed on the events in each 
time-shifted analysis. 

Each candidate event is assigned a p-value that 
measures its significance.  For a candidate event with detection statistic 
$\hat{\rho}_c$, its p-value $p_b$ is the probability that the 
there are one or more coincident noise events (false alarms) that have
a detection statistic value greater than or equal to $\hat{\rho}_c$. 
We calculate p-values under
the null hypothesis that \emph{all} triggers seen are due to noise.  While this
hypothesis might be unrealistic in the presence of loud signals, note that we can
never determine with complete certainty from gravitational-wave data alone that 
any given trigger is due to signal rather than noise.  In order to claim 
detection it is necessary to show that the statistic values of events actually 
obtained in the search are highly improbable under such a null hypothesis: i.e.\ 
that the p-value is very small. 
As demonstrated in~\cite{Capano:2016uif}, including all search triggers in the 
background calculation results in a self-consistent significance, i.e.\ small 
p-values are assigned to random noise candidate events with the expected 
frequency; the procedure also does not adversely affect the efficiency of the 
search compared to other methods which aim to remove likely signal triggers 
before calculating the significance of search events.

Using the distribution of coincident events from the time shifts, we can measure
how many noise background events $n_b$ are louder than a given candidate
event.  We thus determine the function $n_b(\hat{\rho}_c)$ which gives the 
number of background events having a higher detection-statistic value than 
$\hat{\rho}_c$. 
The probability that one or more noise events as loud as a candidate event 
with detection-statistic value $\hat{\rho}^\ast_c$ occurs in the search, 
given the duration of observing time $T$ and the amount of background time 
constructed from the time shifts $T_b$, is 
\begin{equation}
p(\geq 1\;\text{above}\;\hat{\rho}^\ast_c|T, T_b)_0 = 1 - 
 \exp{\left[\frac{-T(1+n_b(\hat{\rho}^\ast_c))}{T_b}\right]}.
\label{eq:pv}
\end{equation}
A detailed derivation of this result is given in \ref{s:appfar}.

To account for the search background noise varying across the target signal
space, candidate and background events are divided into three search classes
based on template length.  The significance of candidate events is measured
against the background from the same class. This binning method prevents loud
background events in one class from suppressing candidates in another class. 
For each candidate event we compute the p-value $p_b^A$ inside the class $A$, 
$p_{b}^A(\hat{\rho}_c) = 
 1 - \exp\left[ -T(1+n_b^A(\hat{\rho}_c)) / T_b \right],
$
where $n_b^A(\hat{\rho}_c)$ is the number of noise background events in the class 
$A$ above the candidate event's re-weighted SNR value $\hat{\rho}_c$. 
To account for having searched multiple classes which could all give rise to 
false alarms, the significance is decreased by a trials factor equal to the 
number of classes $n_C$~\cite{Lyons:1900zz}; the final significance of a 
candidate found in class $A$ is given by
\begin{equation} 
\label{eq:pycbc_slide_fap_trials}
p_{b}(\hat{\rho}_c;A) = 
 1 - \exp\left[\frac{- n_C T(1+n_b^A(\hat{\rho}_c))}{T_b} \right].
\end{equation}
Equivalently, we may consider the quantity $n_C(1 + n_b^A(\hat{\rho}_c))/T_b$ as a
rate of false positive events more significant than such a candidate,
when summed over the search as a whole. 

The observation time used in the search $T_\mathrm{obs}$ is determined by two
considerations. The maximum length of the observation time is set by the 
requirement that the data
quality of the detectors should not change significantly during the search. If
the time-shifts mix data from periods of substantially different data quality
(i.e.\ higher- or lower-than-average trigger rates), the noise background
may not be correctly estimated. The minimum length of the observation time is
set by the false-alarm rate necessary to claim the detection of interesting
signals.  The smallest false-alarm rate that the search 
can measure scales is achieved by performing \emph{every} possible time
shift of the two data streams. While the length of data analyzed in any given
time shift will vary, the total amount of background time analyzed will equal 
$T_b = T ^ 2 / \delta$, where $\delta$ is the time-shift interval.
\footnote{In the case where gaps between consecutive portions of data are 
a multiple of the time-shift interval this is exact.  This is generally the case in our 
analyses. If gaps are not multiples of the time-shift interval, the
total analyzed time will be approximately $T_b = T ^ 2 / \delta$, but there
will be small corrections.}
Thus, the minimum false-alarm rate scales as $\delta / T_\mathrm{obs}^2$.
In a two-detector search using time shifts
of 0.1~seconds, approximately fifteen days of coincident data are sufficient to
measure false-alarm rates of 1 in $2 \times 10^{5}$ years, corresponding
to a significance of 5 $\sigma$.  This is the length of data used in
the observation of GW150914 \cite{Abbott:2016blz}.

Two computational optimizations are applied when calculating the false-alarm
rate of the search. The implementation of the time-shift method used does not
require explicitly checking that triggers are coincident for each time shift
successively.  Instead, the pipeline takes the triggers from a given template
and calculates the offset in their end times from a multiple of the time-shift
interval. It is then possible to quickly find the pairs of triggers that are
within the coincidence window. Furthermore, the number of background triggers
is strongly dependent on the detection-statistic value. For Gaussian noise, the
number of triggers will fall exponentially as a function of $\hat{\rho_c}$.
At low detection-statistic values, it is not necessary to store every noise
event to accurately measure the search false-alarm rate as a function of
detection statistic. Instead, the pipeline is given a threshold on
$\hat{\rho_c}$ and a \emph{decimation} factor $d$.  Below this threshold the
pipeline stores one noise event for every $d$ events, storing the decimation
factor so that the false-alarm rate can be correctly reconstructed. This
saves disk space and makes computation of the false-alarm rate at given values
of the detection statistic faster. 

\section{Comparison to Initial LIGO Pipeline}
\label{s:ihope}

\begin{figure}[tbp]
\begin{center}
\includegraphics[height=8cm]{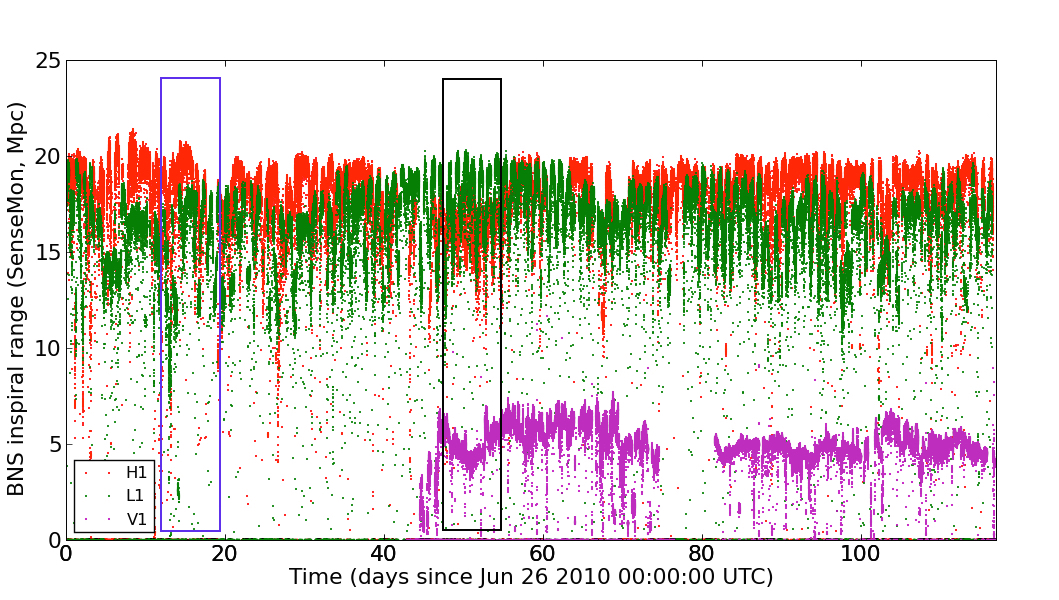} 
\caption{Sensitivity of the gravitational-wave detectors 
for the last part of the sixth science run for LIGO (S6D) and the third VIRGO science run (VSR3). 
The plot shows the volume-weighted average distance at which a 1.4, 1.4 BNS would be
observed with an signal-to-noise ratio of 8 for each detector. 
The two rectangles indicate time intervals used for this study, corresponding
to a week of data from July 2010 and a week of data from August 2010. 
}
\label{img:inspiral-horizon}
\end{center}
\end{figure}

In this section, we compare the PyCBC search pipeline 
to the \emph{ihope} pipeline used in the previous
LIGO and Virgo searches for compact-object binaries \cite{Babak:2012zx}. We
focus on tuning and testing the pipeline to improve
the sensitivity to binary neutron star systems, although we note that use of the pipeline
is not restricted to these sources. This pipeline has been used to search for
binary black holes~\cite{Abbott:2016blz,TheLIGOScientific:2016qqj,TheLIGOScientific:2016pea}, 
binary neutron stars and neutron-star black-hole binaries \cite{O1:BNS_NSBH} in Advanced LIGO data.
Section~\ref{ss:sensitivity}
describes the method that we use to measure the sensitivity of this pipeline
and Section~\ref{ss:results} compares this sensitivity
to that of the ihope pipeline. The comparison is performed using two weeks of data from the sixth LIGO science 
run \cite{Abadie:2011nz, Aasi:2014mqd} shown in Figure~\ref{img:inspiral-horizon}.
We demonstrate that
for binary neutron stars, the PyCBC search can achieve a volume sensitivity
improvement of up to $30\%$ over the ihope pipeline without reducing the
sensitivity to higher mass systems. 

\subsection{Measuring Search Sensitivity}
\label{ss:sensitivity}

As the  metric of search sensitivity, we measure the 
pipeline's \emph{sensitive volume}. This is proportional to the
fraction of sources that the pipeline can detect per unit time at a given false alarm
rate \far{}, given by
\begin{equation}
V(\far) = \int \epsilon(\far; \spatialx, \intparams)\astprior(\spatialx, \intparams)\mathrm{d}\spatialx \mathrm{d}\intparams.
\label{eqn:sensitive_volume}
\end{equation}
Here, \intparams{} are the physical parameters of a signal, \spatialx{} is a spatial co-ordinate,
$\astprior(\spatialx, \intparams)$ is the distribution of signals in the universe, and $\epsilon$ is
the efficiency of the pipeline at detecting signals with parameters
\intparams{} in volume \spatialx{} and with false alarm rate \far{}.

We find the sensitive volume by adding a large number, $N_{I}$, of
simulated signals to the data and measuring the pipeline's ability to identify
them.  If the simulated signals
are drawn from the same distribution as the astrophysical distribution
\astprior{}, then the sensitive volume is
\begin{equation}
V(\far) \approx \frac{1}{N_I} \sum_{i=1}^{N_I} \stepfunc(\far|\far_i) \equiv \left<\stepfunc(\far)\right>\, ,
\label{eqn:vol_monte_carlo}
\end{equation}
where $\far_{i}$ is the false alarm rate associated to simulation $i$ and
$\stepfunc(\far|\far_{i})$ is a step function with $\stepfunc(\far|\far_i) = 1$ if $\far_i \leq \far$
and zero otherwise.  The false alarm rate for a simulation is calculated using
the most significant event within a 1 second window of the arrival time of the
simulated event.  If no event is found within this time, the detection statistic value
is set to zero.

For this comparison, we assume an astrophysical distribution 
\astprior{} in which signals are uniformly distributed in volume, and isotropic 
in sky location and orientation.
\footnote{At small distances, it is necessary to account for discreteness of 
galaxies while, at larger distances, cosmological effects become important.}
Distributing simulations uniform in volume leads to a majority of signals being 
injected at large distances, where almost all will not be detected due to less
than optimal sky position or orientation.  For a fixed number of injections, this
produces large errors in the estimate of $V$.  Instead, we generate signals
distributed uniformly in distance between $r_{\min}$ and $r_{\max}$ chosen 
such that all signals with distance $r < r_{\min}$ will be
found, even at small ($\lesssim 10^{-3}/$yr) false alarm rate thresholds, and
all signals with $r >r_{\max}$ will be missed, even at large ($\gtrsim
10^{2}/$yr) $\far$.  Gravitational waves from more massive systems have 
larger amplitudes and can be detected at greater distances than less massive systems. 
The amplitude of the emitted signal scales, at leading order, with the \emph{chirp mass} of the binary
$\mathcal{M} = (m_1m_2)^{3/5}/(m_1+m_2)^{1/5}$. If a binary with chirp mass
$\mathcal{M}_0$ can be detected at a distance of $r_0$, then a binary with
chirp mass $\mathcal{M}$ can be detected at a distance given approximately by
\begin{equation}
r_i = r_0(\mathcal{M}_i/\mathcal{M}_0)^{5/6}.
\label{eqn:chirp_dist}
\end{equation}
Consequently, we use mass-dependent bounds
$r_{\min,i}$ and $r_{\max,i}$, scaled as in Eq.~(\ref{eqn:chirp_dist}),
for a simulated signal with chirp mass $\mathcal{M}_i$.  The sky locations and
orientations of the simulate population are distributed isotropically.  Since
the distribution of simulated signals differs from the assumed astrophysical 
distribution \astprior{}, we must re-weight the contribution each simulation
makes to the sensitive volume as \cite{Capano:2016dsf}:
\begin{equation}
V(\far) \approx 4\pi\frac{1}{N_I}\sum_{i=1}^{N_I}\left[\frac{1}{3}r^3_{\min,i} + r_i^2\Delta r_i\stepfunc(\far|\far_i)\right] \equiv \left<g(\far)\right>,
\label{eqn:weighted_vol_mc}
\end{equation}
where $\Delta r_i \equiv r_{\max,i} - r_{\min,i}$. The error on this estimate is
\begin{equation}
\delta V = \sqrt{(\left<g(\far)^2\right> - \left<g(\far)\right>^2)/N_I}.
\end{equation}

To quantify the sensitive volume of each pipeline for this study, we select a
compact-object binary population distributed uniformly over component masses with 
$1 \leq m_{1,2}/\mathrm{M}_\odot \leq 7$ and $m_1+m_2 \leq 14\,\mathrm{M}_\odot$. 
The parameters
$\intparams$ are therefore the component masses of the binary $m_1, m_2$ and
the orientation and polarization angle of the binary with respect to the detectors. 
For data from LIGO's sixth science run, we find that $r_{\min} = 0.5\,$Mpc and 
$r_{\max} = 30\,$Mpc are suitable choices for a binary with component masses 
$m_1 = m_2 = 1.4\,\mathrm{M}_\odot$.

\subsection{Relative Search Sensitivity and Computational Cost}
\label{ss:results}

\begin{figure}[tbh]
\begin{center}
	\includegraphics[width=0.8\textwidth]{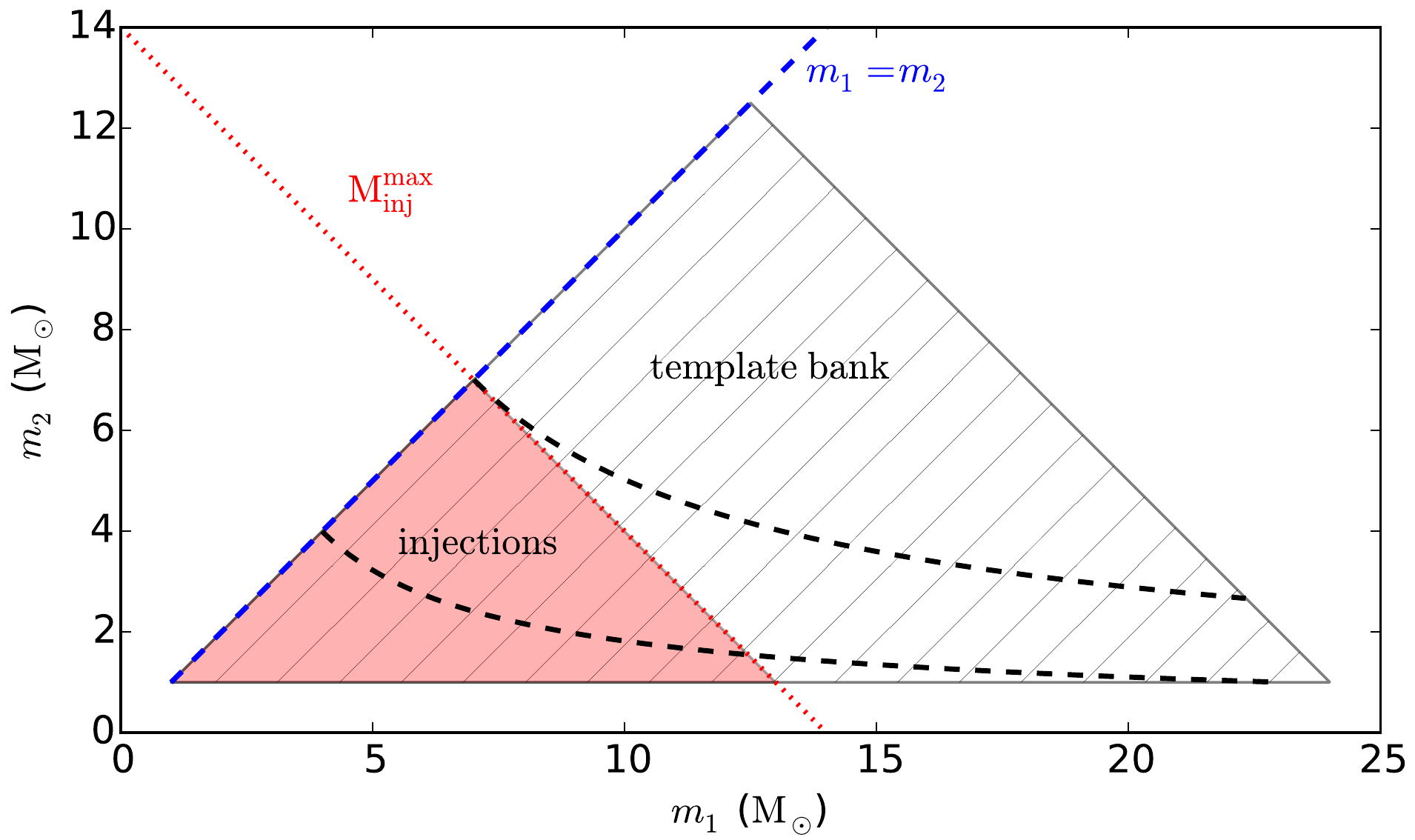} 
\caption{Mass-ranges for software injection, shown in the $m_1-m_2$
mass-plane. The template bank
used to search for these injections is indicated by hatched regions and the
injection set by the red shaded region. The black dashed lines show chirp
masses of $3.48\,\mathrm{M}_\odot$ and $6.1\,\mathrm{M}_\odot$, the boundaries 
between the mass bins used. Triggers from templates with chirp masses larger 
than $6.1\,\mathrm{M}_\odot$ are discarded in post-processing.
}
\label{Inj-massrange}
\end{center}
\end{figure}

Both the PyCBC and ihope~\cite{Babak:2012zx} pipelines were used to analyze two
weeks of data from LIGO's sixth science runs and the sensitive volume of each
search, as well as the computational cost, was compared.  Both searches 
use a template bank designed to search for compact-object binaries with component masses between
$1$ and $24\,\mathrm{M}_\odot$, total mass $m_1 + m_2 \le 25\,\mathrm{M}_\odot$ and zero spins, as shown
in Fig.~\ref{Inj-massrange}. In the literature it is generally assumed that the noise power spectral
density is invariant, which is not the case for real data. In searches of
initial LIGO and Virgo data, the time-dependence of the detector PSDs was
addressed by recomputing the template bank on regular intervals of around an
hour~\cite{Babak:2012zx}. Since PyCBC's templates include both mass and spin
parameters, regenerating the bank so often is not computationally feasible; 
instead, the use of a single, pre-generated bank is required, as described in
Section~\ref{s:bank}.

The PyCBC analysis
was configured as described in detail in Sections \ref{s:overview} and \ref{s:description}.  
The chi-square test uses $p=100$ bins for templates with a chirp mass 
$\mathcal{M} \le 1.74\,\mathrm{M}_\odot$ and $p=16$ for higher masses.  To estimate the noise
background of the search, an $0.2$\,s time-shift interval is used, and events are divided
into two classes: one class contains triggers with $\mathcal{M} \le 1.74\,\mathrm{M}_\odot$, a 
second triggers with $1.74 < \mathcal{M}/\mathrm{M}_\odot \le 6.1$, and templates outside these 
bins are ignored in the search. 

The ihope analysis was configured as for the S6-VSR2/3 search for low-mass
binaries~\cite{Abadie:2011nz}.  To reduce computational cost, the ihope pipeline
has \textit{two stages} of matched filtering, where the computationally costly chi-square 
test is performed at the second stage on only those triggers found in coincidence 
(and time-shift coincidence).  When large numbers of time shifts are performed, the
majority of triggers appear in coincidence in at least one time shift and there is
little benefit to the two-stage analysis.  Thus, to investigate sensitivity at low
false alarm rates, we also run the pipeline with a \textit{single stage} of matched
filtering where the chi-squared test is performed on every trigger.

\begin{figure}[tbp]
\begin{center}
\includegraphics[width=0.49\linewidth]{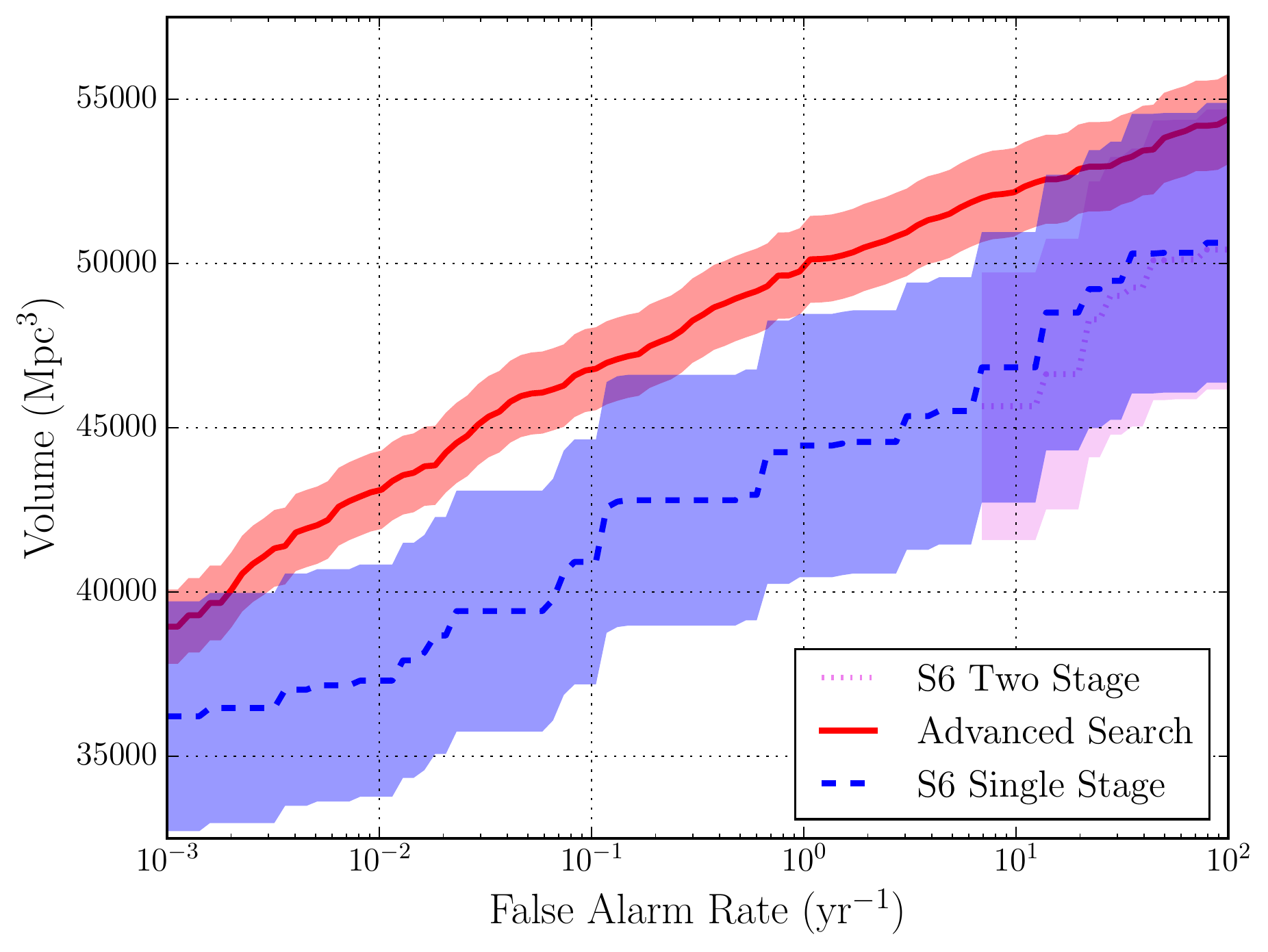}
\includegraphics[width=0.49\linewidth]{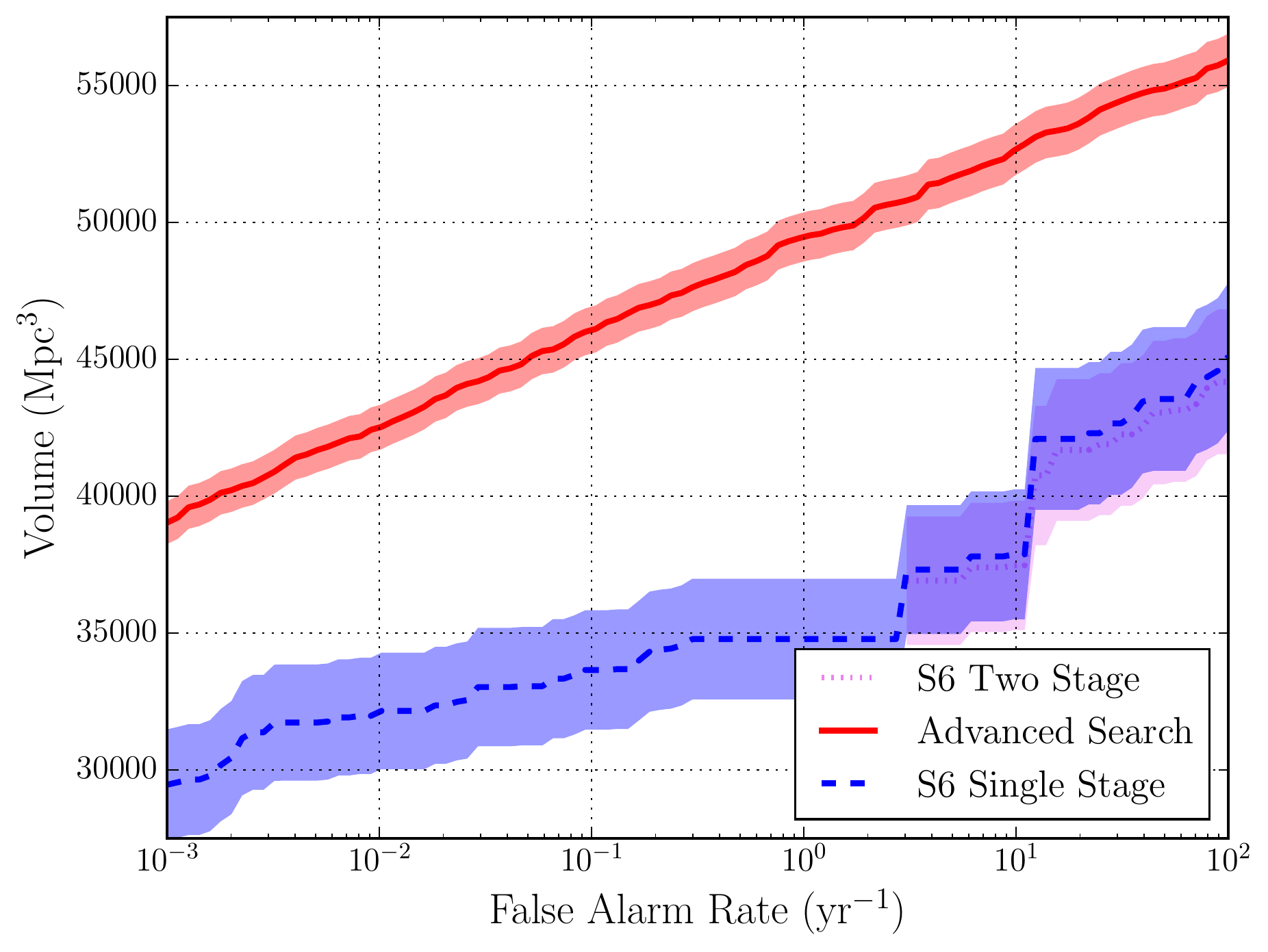}
\caption{The sensitive volume for low-mass binary systems, 
chirp mass $\mathcal{M}\leq 3.48 \mathrm{M}_{\odot}$, 
of the PyCBC and ihope searches as a function of the 
false alarm rate threshold.  The left plot 
shows results from a week of data from July 2010 while the right plot 
uses data from August 2010. The 
PyCBC search (red, solid line) is more sensitive over a broad range of
false alarm rates than the ihope single stage (blue, dashed line) 
and ihope two-stage (pink dotted line) searches.  The error bars for the
ihope analyses are larger than for PyCBC as fewer injections were performed.
Since it uses only 100 time shifts, the two-stage ihope analysis can only determine 
false alarm rates as low as six per year in one week of data.}
\label{fig:sensitivity_low-mass}
\end{center}
\end{figure}

\begin{figure}[tbp]
\begin{center}
\includegraphics[width=0.49\linewidth]{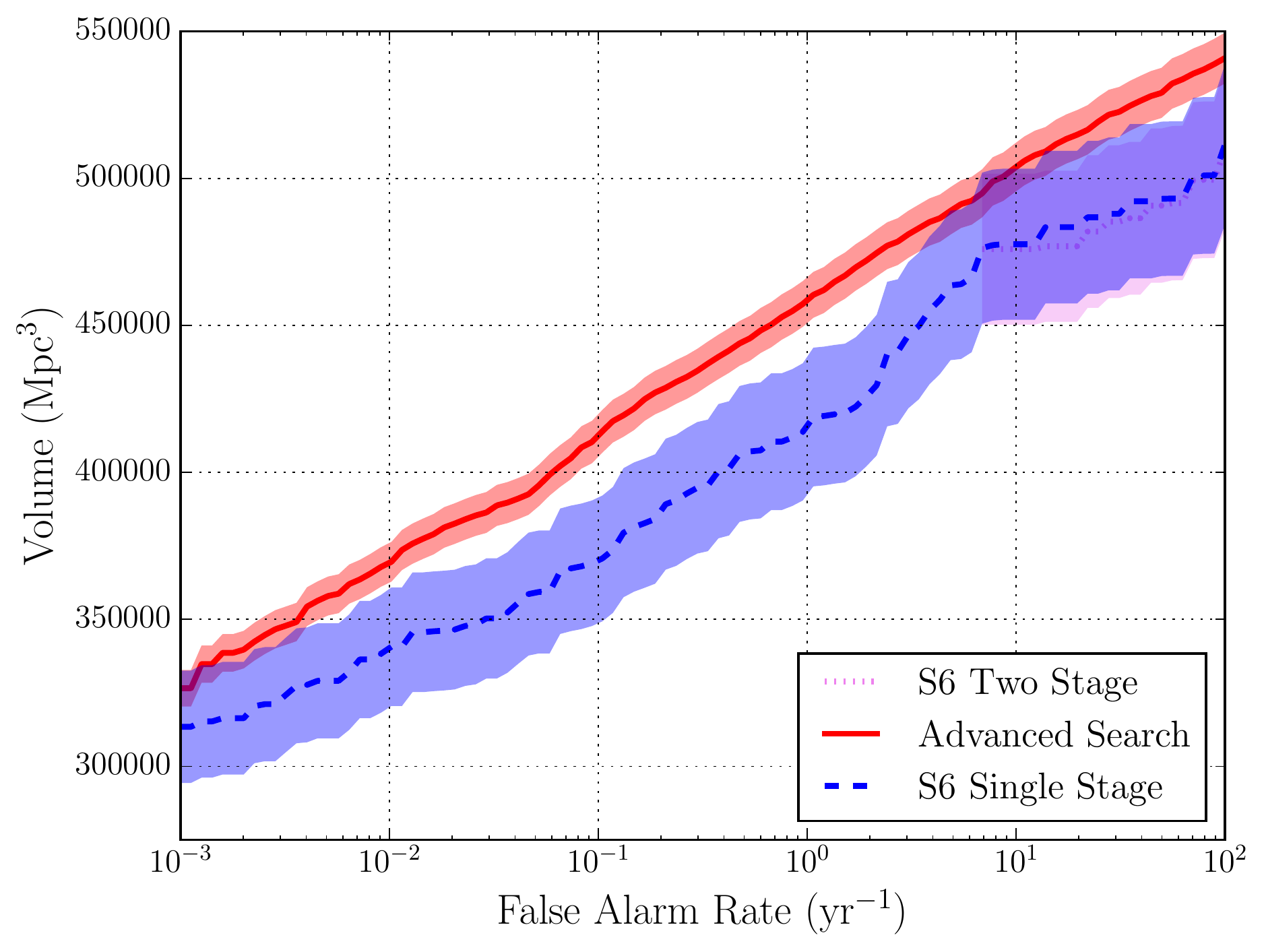}
\includegraphics[width=0.49\linewidth]{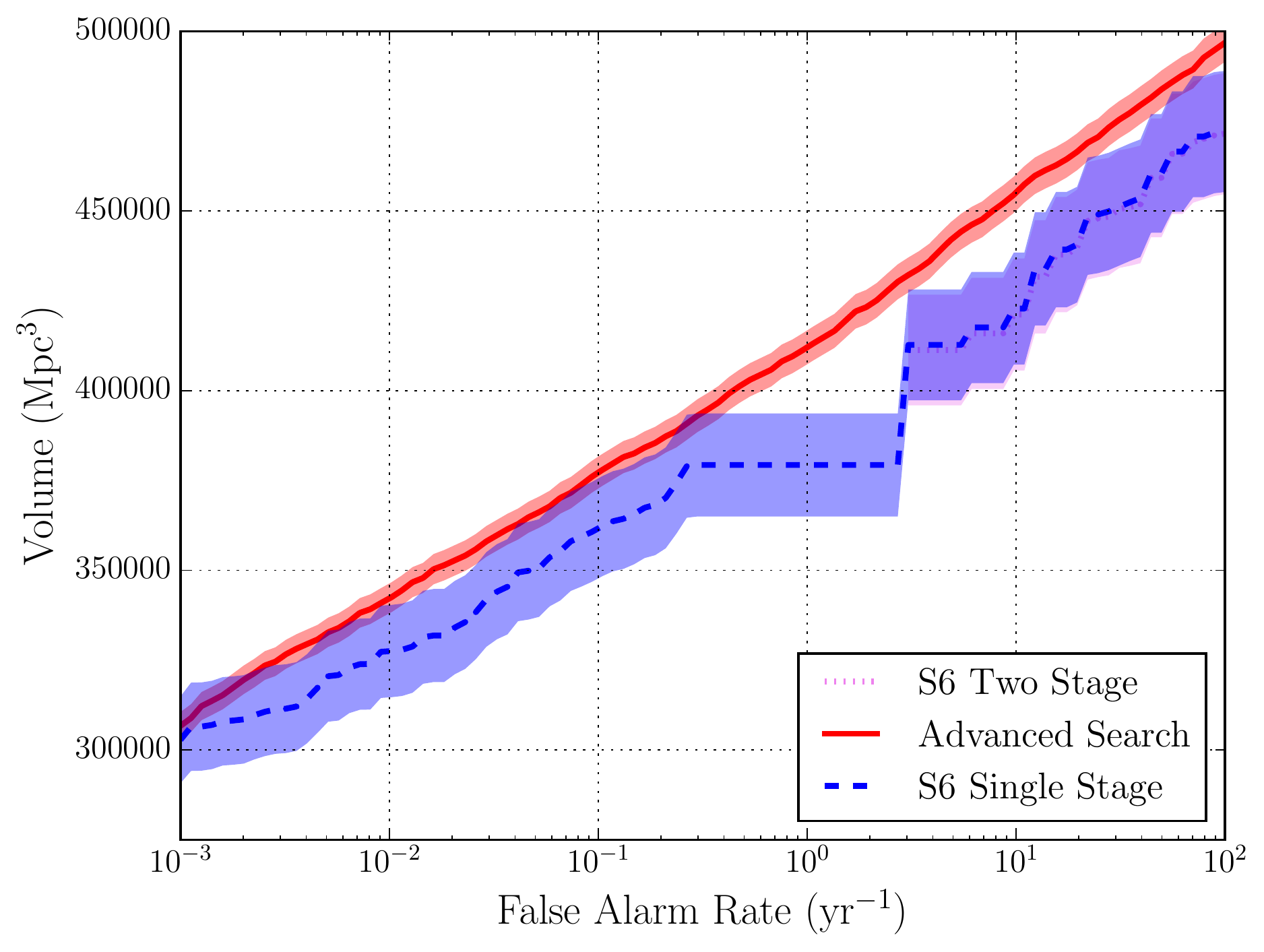}
\caption{The sensitive volume for high-mass binary systems, 
chirp mass $\mathcal{M} > 3.48 \mathrm{M}_{\odot}$,
of the PyCBC and ihope searches as a function of the 
false alarm rate threshold.  The left plot 
shows results from a week of data from July 2010 while the right plot 
uses data from August 2010.  The 
PyCBC search (red, solid line) is more sensitive over a broad range of
false alarm rates than the ihope single stage (blue, dashed line) 
and ihope two-stage (pink dotted line) searches.  The error bars for the
ihope analyses are larger than for PyCBC as fewer injections were performed.
Since it uses only 100 time shifts, the two-stage ihope analysis can only determine 
false alarm rates as low as six per year in one week of data.}
\label{fig:sensitivity_high-mass}
\end{center}
\end{figure}

Figures~\ref{fig:sensitivity_low-mass} and \ref{fig:sensitivity_high-mass} compare the
sensitivity of the two pipelines, using a week of data from July 2010 and a second week
from August 2010. The data from July had larger fluctuations in the overall sensitivity, 
but relatively few noise transients while the August data had a consistent range but more 
glitches~\cite{Aasi:2014mqd}. 
Using these two different weeks allows us to test the pipelines under various conditions.
Figure~\ref{fig:sensitivity_low-mass} restricts to low mass binary coalescence signals 
($\mathcal{M}\leq 3.48 \mathrm{M}_{\odot}$) while Figure~\ref{fig:sensitivity_high-mass}
displays higher masses.  The figures show that the sensitive
volume of the new pipeline is greater than or equal to that of the ihope
pipeline for both subsets of signals in both weeks of data.

At a false-alarm rate of 1 in 100 years, the PyCBC pipeline affords a sensitivity improvement 
for low-mass systems of 20\% for the first week of data and 30\% for the second week of data.
The matched-filter SNR of simulated signals does not
change significantly between the old and new pipelines. 
Furthermore, the chi-squared value, and hence re-weighted SNR, is also similar for both pipelines. 
The improvement in sensitivity of the PyCBC analysis is due to the reduction in the
noise background of the search.  Approximately
$10\%$ of the improvement in sensitive volume comes from using a fixed
template bank with the exact-match coincidence test. 
The remaining improvement is due to reduction in the noise background because of the
increased number of chi-squared bins used in the low-mass region of the search.  Both
of these improvements allow us to decrease the number of high significance events
caused by noise transients.
The improvement is most apparent in the August data  as the rate of non-Gaussian noise 
transients in that week is larger than in the July data.
Consequently, the new pipeline's reduction of the noise
background has a larger effect on the sensitivity for the August 2010 data.  
For high-mass systems, the PyCBC pipeline is marginally more sensitive than 
the ihope pipeline, with a 5-10\% improvement at a false-alarm rate of 1 in 100 years. 
The improvement is less pronounced as the number of chi-squared bins was unchanged
at higher masses.  A discussion of tuning the pipeline to 
binary black hole systems can be found in Ref.~\cite{TheLIGOScientific:2016qqj}. 

\begin{table}[tbh]
\begin{center}
\begin{tabular}{|c|c|c|c|c|}
\hline 									
Job Type & Two-Stage \texttt{ihope} &	Single-Stage \texttt{ihope} & \texttt{PyCBC} \\\hline
Template Bank Generation    &    13.3  &  13.3 &  4.7	\\ \hline
Matched filtering and $\chi^2$  &	515.4	&  1431 &  515.5	\\ \hline
Second Template Bank      &	0.1	&  - &  -	\\ \hline
Coincidence Test        &	0.3	&  8.3 &  9.9	\\ \hline
Total              &	529.1	&  1453 &  530.0	\\ \hline
\end{tabular}		
\end{center}
\caption{The computational costs of different parts of the
single-stage and two-stage ihope search pipelines, and the new PyCBC pipeline. The costs are given in CPU days.}
\label{t:cost}
\end{table}

Table~\ref{t:cost} shows the computational cost of the pipelines.  The cost of the 
PyCBC search is comparable to that of the two-stage ihope search, and about one
third of the cost of the single-stage ihope search.  However, the
two-stage search is unable to compute false-alarm rates low enough to support
a detection claim and is therefore unsuitable for analysis of advanced detector data.

\section{Conclusions}
\label{s:conclusions}

This paper presents a new pipeline to search for gravitational
waves from compact binaries developed in the PyCBC framework. The pipeline includes
several new developments, including:
(i) gating to mitigate the effect of loud, non-Gaussian noise transients in
the input data; (ii) the ability to use a single, fixed template bank for the
entire analysis created using the harmonic mean of a large number of noise
spectral density estimates; (iii) decoupling of the matched filtering and noise
power spectral density estimation to improve the performance of the matched
filter; (iv) a simpler, more efficient algorithm for identifying and
clustering peaks in the matched-filter SNR time-series; (v) a 
computationally-efficient implementation of the chi-squared veto; 
(vi) the ability to set the number of bins in the
chi-squared veto as a function of the parameters of the template; (vii) a
stricter coincidence test that requires that a trigger is found with exactly
the same template in both detectors; and (viii) improvements to the time-shift
method used to measure the false-alarm rate of the search and assign a p-value
to candidate events.  The new pipeline provides an improvement in 
sensitivity over the ihope pipeline ~\cite{Babak:2012zx} used in searches of initial 
LIGO and Virgo data \cite{Abadie:2011nz}, with up to 30\% increase in sensitivity to 
binary neutron star sources.
The PyCBC pipeline described here was used in the LIGO Scientific Collaboration and
the Virgo Collaboration's search for binary black holes in the first science
run of Advanced LIGO~\cite{TheLIGOScientific:2016qqj, TheLIGOScientific:2016pea, O1:BNS_NSBH},
including the identification of the binary black hole signals GW150914~\cite{Abbott:2016blz}
and GW151226~\cite{Abbott:2016nmj}.

The PyCBC framework provides a modular set of tools and algorithms that
can be used to create flexible, sensitive gravitational wave search pipelines.  For
comparison presented in Section~\ref{s:ihope}, we deliberately configured the pipeline 
to mimic the ihope analysis as closely as possible.  
However, the PyCBC infrastructure can be configured in many different ways.  For 
example, the mass range can be significantly extended \cite{TheLIGOScientific:2016qqj, 
TheLIGOScientific:2016pea}, and the choice of template waveform can be varied across
the mass space: e.g. Refs.~\cite{Abbott:2016blz,TheLIGOScientific:2016qqj} use
post-Newtonian templates for system with a total mass less that $4\,\mathrm{M}_\odot$
and templates from the Effective One Body family for higher-mass
systems~\cite{Buonanno:1998gg,Taracchini:2013rva,Purrer:2015tud}. 
In principle, it is also possible to
use the methods described here to implement matched filtering to detect
gravitational waves from modeled transient systems other than compact-object
binaries.

Further improvements to the astrophysical sensitivity of the PyCBC search pipeline are
ongoing. These include the extension of the pipeline to multiple
detectors, upgrades to the detection statistic to include signal amplitude
and phase consistency between detectors and the incorporation of methods to 
better measure the variation of the noise event distribution across the
template bank. Ongoing advancements to the computational efficiency include
investigating changes to the FFT block size and input data sample rate and
the implementation of hierarchical search methods in the matched filter. The FFT
engine used in this pipeline can be efficiently used on graphics
processing units (GPUs) and work is ongoing to develop a fast, efficient GPU
implementation of this pipeline. In the longer term, the pipeline will be extended
to search for binaries that exhibit spin-induced precession~\cite{Harry:2016ijz}, 
include the effects of higher-order multipoles~\cite{Capano:2013raa}, perform a
coherent search using a global network of detectors~\cite{Macleod:2015jsa} and
target the search to look for gravitational waves emitted at the time of short gamma-ray 
bursts or other astronomical transients~\cite{Harry:2010fr, Williamson:2014wma}.
Results from these studies will be reported
in future publications.

\ack
CMB, DAB, AHN, and SAU acknowledge support from NSF awards PHY-0847611 and
PHY-1404395, ACI-1443037 and a Research Corporation for Science Advancement Cottrell
Scholar Award. IWH, PRS and MW acknowledge support from NSF awards PHY-0854812 and
PHY-1205835. JLW acknowledges support from NSF award PHY-1506254. MSK, IWH, CDC,
TD, TDC, JLW, and DK acknowledge and thank the support of the Max-Planck-Gesellschaft. 
MSK and HPP acknowledge support from NSERC of Canada, the Ontario Early
Researcher Awards Program, the Canada Research Chairs Program, and 
the Canadian Institute for Advanced Research. 
MSK and HPP are grateful for hospitality of the Max-Planck-Institut for
Gravitationsphysik.
SF acknowledges support from the Royal Society and STFC award ST/L000962/1.
Computations used in this analysis were performed on the
Syracuse University SUGAR cluster, supported by NSF awards PHY-1040231 and
PHY-1104371, ACI-1341006 and by Syracuse University ITS, as well as the ATLAS cluster
supported by the Max-Planck-Gesellschaft. We thank Soumi De and Steven Reyes
for their comments on this paper.

\appendix

\section{Calculation of Event Significance}
\label{s:appfar}

In this Appendix we give the derivation of Equation~(\ref{eq:pv}) which 
gives the probability that one or more events with higher detection 
statistic value than a given threshold $\hat{\rho}^\ast_c$ would occur 
due to random coincidences of noise triggers over the search 
duration.\footnote{A less detailed derivation was presented in~\cite{Capano:2016uif}.}

For a randomly chosen candidate event which is assumed to be due to noise, 
the probability that there are $n^\ast$ or fewer background events with 
a higher value of $\hat{\rho}^\ast_c$ 
is given by
\begin{equation} 
 p(n_b \leq n^\ast|N_e=1,N_b)_0 = \frac{1 + n^\ast}{1 + N_b},
\end{equation}
where $N_b$ is the total number of background events and $N_e$ is the number
of candidate events under consideration.  This result follows by comparing a 
random (noise) candidate event's $\hat{\rho}_c$ value to an ordered list of background 
events from the time shifts. The candidate value can lie above all background 
events, below all background events or lie in between two background events. 
There are $N_b + 1$ places where a given candidate event can lie when ranked 
against the list of background events. If the candidate event is due to noise 
then its detection statistic value is drawn from the same distribution as the 
time-shifted background events, therefore it is equally likely to occupy any 
one of these $N_b + 1$ positions. 

Now, since higher $\hat{\rho}_c$ values correspond to smaller $n_b$ values, the 
probability that one random coincident noise event lies above a threshold 
$\hat{\rho}^\ast_c$ is
\begin{equation}
 p(\hat{\rho}_c \geq \hat{\rho}^\ast_c|N_e=1, N_b)_0 = 
 \frac{1 + n_b(\hat{\rho}^\ast_c)}{1 + N_b}.
\end{equation}
The same count of louder background events $n_b$ might be obtained for a 
slightly smaller value of $\hat{\rho}_c$, given that background event values 
are not infinitely finely spaced, thus the condition $\hat{\rho}_c \geq 
\hat{\rho}^\ast_c$ is more restrictive than $n_b \leq n_b(\hat{\rho}^\ast_c)$, 
and so strictly this equation should have a $\leq$ sign; however we will neglect 
this subtlety in what follows. 
We wish to find the probability that one or more events out of the $N_e$ 
coincident events in the search is a noise event (false alarm) at or above 
the threshold $\hat{\rho}^\ast_c$.  This is given by the complement of the 
probability that all events do \emph{not} lie above this threshold, which for one 
event is given by $1-\left[1 + n_b(\hat{\rho}^\ast_c)\right]/\left[1 + N_b\right]$.
The probability that this is true for all candidate events follows by 
multiplying the individual probabilities for each of the $N_e$ events: 
\begin{equation}
 p(\text{none above}\;\hat{\rho}^\ast_c|N_e, N_b)_0 = 
 \left(1 - \frac{1+n_b(\hat{\rho}^\ast_c)}{1 + N_b} \right)^{N_e}.
\end{equation}
This step requires that candidate events are independent, which is achieved,
to good approximation, by the final event clustering described in Section~\ref{s:far}. 
Then the probability that at least one 
out of $N_e$ clustered candidate events \emph{is} louder than a threshold 
$\hat{\rho}^\ast_c$ (if all candidate events are due to noise) is 
\begin{equation} 
\label{eq:rate}
 p(\geq 1\;\text{above}\;\hat{\rho}^\ast_c|N_e, N_b)_0 = 
 1 - \left(1 - \frac{1+n_b(\hat{\rho}^\ast_c)}{1 + N_b} \right)^{N_e}.
\end{equation}
In what follows we will approximate $1+N_b \simeq N_b$, which is a 
negligible correction given that $N_b$ is very large.

Since we do not know $N_b$ and $N_e$ before performing the analysis, we 
should treat these event counts as stochastic variables.  However, since 
there is such a large number of background events, the statistical 
uncertainty on $N_b$ is negligibly small. We expect the rate of background 
noise events for the duration of background time $T_b$ analyzed to be 
equal to the rate of coincident events over the duration of foreground time 
analyzed, $T$:
\begin{equation}
\frac{N_e}{T} \approx \frac{N_b}{T_b}.
\end{equation}
Under the assumption that the candidate events are all due to noise, we can 
model $N_e$ as a Poisson distribution with a mean of $\langle N_e \rangle_0 
= (T/T_b)N_b$.

We wish to know the significance of obtaining a candidate with a given 
$\hat{\rho}^\ast_c$ without restricting to any specific number of 
coincident events, therefore we marginalize Eq.~(\ref{eq:rate}) over $N_e$ 
to obtain
\begin{equation}
\label{eq:sum}
 p(\geq 1\;\text{above}\;\hat{\rho}^\ast_c|N_b)_0 = \sum_{N_e}p(\geq
 1\;\text{above}\;\hat{\rho}^\ast_c|N_e, N_b)_0 p(N_e|N_b). 
\end{equation}
Given that coincident noise events are approximated by a Poisson process 
that we measure using the time-shifted background events, we can find the 
unknown probability $p(N_e|N_b)$. Letting the Poisson rate of coincident 
noise events be $\mu = (N_b T)/T_b$, then the probability of obtaining $N_e$ 
events is
\begin{equation}
p(N_e|N_b) \equiv p(N_e|\mu) = \mu^{N_e}\frac{\exp(-\mu)}{N_e!}.
\end{equation}
Substituting into Eq.~(\ref{eq:sum}) we obtain
\begin{equation}
p(\geq 1\;\text{above}\;\hat{\rho}^\ast_c|N_b)_0 = 
 \sum_{N_e}\left\{1 - \left[1 - \frac{1+n_b(\hat{\rho}^\ast_c)}{N_b} \right]^{N_e}\right\}
 \mu^{N_e}\frac{\exp(-\mu)}{N_e!}.
\end{equation}
Since the sum of a Poisson distribution over all possible event counts
$\sum_{N_e}\mu^{N_e}\frac{\exp(-\mu)}{N_e!}$ is unity, this simplifies to
\begin{equation}
\label{eq:blah}
p(\geq 1\;\text{above}\;\hat{\rho}^\ast_c|N_b)_0 = 
 1 - \sum_{N_e} \left(\mu\left[1 - \frac{1+n_b(\hat{\rho}^\ast_c)}{N_b} \right]\right)^{N_e}
 \frac{\exp(-\mu)}{N_e!}.
\end{equation}
Next, we multiply inside the summation by 
$\exp[-\mu(1-\frac{1+n_b}{N_b})] \exp[\mu(1-\frac{1+n_b}{N_b})]$ (equivalent
to multiplication by unity), to
rewrite Eq.~(\ref{eq:blah}) as:
\begin{multline}
p(\geq 1\;\text{above}\;\hat{\rho}^\ast_c|N_b)_0 = 
 1 - \sum_{N_e}\left(\mu\left[1 - \frac{1+n_b(\hat{\rho}^\ast_c)}{N_b} \right]\right)^{N_e} \\
\times
 \frac{\exp\left[-\mu(1-(\frac{1+n_b(\hat{\rho}^\ast_c)}{N_b}))\right]}{N_e!}
 \exp \left[\frac{-\mu(1+n_b(\hat{\rho}^\ast_c))}{N_b}\right].
\end{multline}
Setting $\hat{\mu}$ equal to $\mu[1-(1+n_b)/N_b]$, we again identify a sum 
of the Poisson probability over all possible counts:
\begin{equation}
p(\geq 1\;\text{above}\;\hat{\rho}^\ast_c|N_b)_0 = 
 1 - \sum_{N_e}\hat{\mu}^{N_e}\frac{\exp(-\hat{\mu})}{N_e!}
 \exp \left[\frac{-\mu(1+n_b(\hat{\rho}^\ast_c))}{N_b}\right],
\end{equation}
All but the last term in the sum total to one and we can re-write this
using the Poisson rate $\mu=T(N_b/T_b)$, giving 
\begin{equation}
p(\geq 1\;\text{above}\;\hat{\rho}^\ast_c|T, T_b)_0 = 1 - 
 \exp{\left[\frac{-T(1+n_b(\hat{\rho}^\ast_c))}{T_b}\right]},
\end{equation}
which reproduces Eq.~(\ref{eq:pv}) in the main text. 

\section*{References}

\providecommand{\newblock}{}


\begin{thebibliography}{10}
\expandafter\ifx\csname url\endcsname\relax
  \def\url#1{{\tt #1}}\fi
\expandafter\ifx\csname urlprefix\endcsname\relax\def\urlprefix{URL }\fi
\providecommand{\eprint}[2][]{\url{#2}}

\bibitem{Abbott:2016blz}
Abbott B~P {\em et~al.\/} (Virgo, LIGO Scientific) 2016 {\em Phys. Rev.
  Lett.\/} {\bf 116} 061102 (\textit{Preprint} \eprint{1602.03837})

\bibitem{Abbott:2016nmj}
Abbott B~P {\em et~al.\/} (Virgo, LIGO Scientific) 2016 {\em Phys. Rev.
  Lett.\/} {\bf 116} 241103 (\textit{Preprint} \eprint{1606.04855})

\bibitem{TheLIGOScientific:2016pea}
Abbott B~P {\em et~al.\/} (Virgo, LIGO Scientific) 2016  (\textit{Preprint}
  \eprint{1606.04856})

\bibitem{Abbott:2007kv}
Abbott B {\em et~al.\/} (LIGO Scientific) 2009 {\em Rept.Prog.Phys.\/} {\bf 72}
  076901 (\textit{Preprint} \eprint{0711.3041})

\bibitem{TheLIGOScientific:2016agk}
Abbott B~P {\em et~al.\/} (Virgo, LIGO Scientific) 2016 {\em Phys. Rev.
  Lett.\/} {\bf 116} 131103 (\textit{Preprint} \eprint{1602.03838})

\bibitem{Accadia:2012zzb}
Accadia T {\em et~al.\/} (VIRGO) 2012 {\em JINST\/} {\bf 7} P03012

\bibitem{Acernese:2015gua}
Acernese F (Virgo) 2015 {\em J.Phys.Conf.Ser.\/} {\bf 610} 012014

\bibitem{Akutsu:2015hua}
Akutsu T (KAGRA) 2015 {\em J.Phys.Conf.Ser.\/} {\bf 610} 012016

\bibitem{Abbott:2016nhf}
Abbott B~P {\em et~al.\/} (Virgo, LIGO Scientific) 2016  (\textit{Preprint}
  \eprint{1602.03842})

\bibitem{Abadie:2010cf}
Abadie J {\em et~al.\/} (LIGO Scientific, VIRGO) 2010 {\em Class.Quant.Grav.\/}
  {\bf 27} 173001 (\textit{Preprint} \eprint{1003.2480})

\bibitem{Aasi:2013wya}
Aasi J {\em et~al.\/} (LIGO Scientific, VIRGO) 2013  (\textit{Preprint}
  \eprint{1304.0670})

\bibitem{O1:BNS_NSBH}
Abbott B~P {\em et~al.\/} (Virgo, LIGO Scientific) 2016  (\textit{Preprint}
  \eprint{1607.07456})

\bibitem{pycbc-github}
Nitz A~H, Harry I~W, Willis J~L, Biwer C~M, Brown D~A, Pekowsky L~P, Dal~Canton
  T, Williamson A~R, Dent T, Capano C~D, Massinger T~T, Lenon A~K, Nielsen A
  and Cabero M 2016 {PyCBC Software} \url{https://github.com/ligo-cbc/pycbc}

\bibitem{Brown:2004pv}
Brown D~A {\em et~al.\/} 2004 {\em Class. Quant. Grav.\/} {\bf 21} S1625--S1633
  (\textit{Preprint} \eprint{arXiv:0705.1572 [gr-qc]})

\bibitem{Brown:2005zs}
Brown D~A (LIGO) 2005 {\em Class. Quant. Grav.\/} {\bf 22} S1097--S1108
  (\textit{Preprint} \eprint{gr-qc/0505102})

\bibitem{Allen:2004gu}
Allen B 2005 {\em Phys.Rev.\/} {\bf D71} 062001 (\textit{Preprint}
  \eprint{gr-qc/0405045})

\bibitem{Allen:2005fk}
Allen B, Anderson W~G, Brady P~R, Brown D~A and Creighton J~D 2012 {\em
  Phys.Rev.\/} {\bf D85} 122006 (\textit{Preprint} \eprint{gr-qc/0509116})

\bibitem{Babak:2012zx}
Babak S, Biswas R, Brady P, Brown D, Cannon K {\em et~al.\/} 2013 {\em
  Phys.Rev.\/} {\bf D87} 024033 (\textit{Preprint} \eprint{1208.3491})

\bibitem{Abbott:2003pj}
Abbott B {\em et~al.\/} (LIGO Scientific) 2004 {\em Phys. Rev.\/} {\bf D69}
  122001 (\textit{Preprint} \eprint{gr-qc/0308069})

\bibitem{Abbott:2005pe}
Abbott B {\em et~al.\/} (LIGO Scientific) 2005 {\em Phys. Rev.\/} {\bf D72}
  082001 (\textit{Preprint} \eprint{gr-qc/0505041})

\bibitem{Abbott:2005kq}
Abbott B {\em et~al.\/} (LIGO Scientific) 2006 {\em Phys. Rev.\/} {\bf D73}
  062001 (\textit{Preprint} \eprint{gr-qc/0509129})

\bibitem{Abbott:2007xi}
Abbott B {\em et~al.\/} (LIGO Scientific) 2008 {\em Phys. Rev.\/} {\bf D77}
  062002 (\textit{Preprint} \eprint{0704.3368})

\bibitem{Abbott:2007ai}
Abbott B {\em et~al.\/} (LIGO Scientific) 2008 {\em Phys. Rev.\/} {\bf D78}
  042002 (\textit{Preprint} \eprint{0712.2050})

\bibitem{Abbott:2009tt}
Abbott B~P {\em et~al.\/} (LIGO Scientific) 2009 {\em Phys. Rev.\/} {\bf D79}
  122001 (\textit{Preprint} \eprint{0901.0302})

\bibitem{Abbott:2009qj}
Abbott B~P {\em et~al.\/} (LIGO Scientific) 2009 {\em Phys. Rev.\/} {\bf D80}
  047101 (\textit{Preprint} \eprint{0905.3710})

\bibitem{Abadie:2010yba}
Abadie J {\em et~al.\/} (LIGO Scientific) 2010 {\em Phys. Rev.\/} {\bf D82}
  102001 (\textit{Preprint} \eprint{1005.4655})

\bibitem{Abadie:2011nz}
Abadie J {\em et~al.\/} (The LIGO Scientific) 2012 {\em Phys.Rev.\/} {\bf D85}
  082002

\bibitem{TheLIGOScientific:2016qqj}
Abbott B~P {\em et~al.\/} (Virgo, LIGO Scientific) 2016  (\textit{Preprint}
  \eprint{1602.03839})

\bibitem{Blanchet:2013haa}
Blanchet L 2014 {\em Living Rev.Rel.\/} {\bf 17} 2 (\textit{Preprint}
  \eprint{1310.1528})

\bibitem{Taracchini:2013rva}
Taracchini A, Buonanno A, Pan Y, Hinderer T, Boyle M {\em et~al.\/} 2014 {\em
  Phys.Rev.\/} {\bf D89} 061502 (\textit{Preprint} \eprint{1311.2544})

\bibitem{wainstein:1962}
Wainstein L~A and Zubakov V~D 1962 {\em Extraction of signals from noise\/}
  (Englewood Cliffs, NJ: Prentice-Hall)

\bibitem{Cutler:1992tc}
Cutler C, Apostolatos T~A, Bildsten L, Finn L~S, Flanagan E~E {\em et~al.\/}
  1993 {\em Phys.Rev.Lett.\/} {\bf 70} 2984--2987 (\textit{Preprint}
  \eprint{astro-ph/9208005})

\bibitem{Aasi:2014mqd}
Aasi J {\em et~al.\/} (LIGO Scientific, VIRGO) 2015 {\em Class.Quant.Grav.\/}
  {\bf 32} 115012 (\textit{Preprint} \eprint{1410.7764})

\bibitem{GW150914-DETCHAR}
Abbott B~P, Abbott R, Abbott T~D {\em et~al.\/} (LIGO Scientific Collaboration,
  Virgo Collaboration) 2016  {\bf 33} 134001 (\textit{Preprint}
  \eprint{1602.03844})

\bibitem{Creighton:1999qw}
Creighton J~D~E 1999 {\em Phys. Rev.\/} {\bf D60} 021101 (\textit{Preprint}
  \eprint{gr-qc/9901075})

\bibitem{Siemens:2004pr}
Siemens X, Allen B, Creighton J, Hewitson M and Landry M 2004 {\em Class.
  Quant. Grav.\/} {\bf 21} S1723--S1736 (\textit{Preprint}
  \eprint{gr-qc/0405070})

\bibitem{Abadie:2010px}
Abadie J {\em et~al.\/} (LIGO Scientific) 2010 {\em Nucl.Instrum.Meth.\/} {\bf
  A624} 223--240 (\textit{Preprint} \eprint{1007.3973})

\bibitem{Nuttall:2015dqa}
Nuttall L {\em et~al.\/} 2015 {\em Class. Quant. Grav.\/} {\bf 32} 245005
  (\textit{Preprint} \eprint{1508.07316})

\bibitem{Slutsky:2010ff}
Slutsky J {\em et~al.\/} 2010 {\em Class. Quant. Grav.\/} {\bf 27} 165023
  (\textit{Preprint} \eprint{1004.0998})

\bibitem{Blackburn:2008ah}
Blackburn L {\em et~al.\/} 2008 {\em Class. Quant. Grav.\/} {\bf 25} 184004
  [Class. Quant. Grav.25,184004(2008)] (\textit{Preprint} \eprint{0804.0800})

\bibitem{Aasi:2012wd}
Aasi J {\em et~al.\/} (VIRGO Collaboration) 2012 {\em {Class. Quant. Grav.}\/}
  {\bf 29} 155002 (\textit{Preprint} \eprint{1203.5613})

\bibitem{Sathyaprakash:1991mt}
Sathyaprakash B and Dhurandhar S 1991 {\em Phys.Rev.\/} {\bf D44} 3819--3834

\bibitem{Dhurandhar:1992mw}
Dhurandhar S and Sathyaprakash B 1994 {\em Phys.Rev.\/} {\bf D49} 1707--1722

\bibitem{Owen:1995tm}
Owen B~J 1996 {\em Phys.Rev.\/} {\bf D53} 6749--6761 (\textit{Preprint}
  \eprint{gr-qc/9511032})

\bibitem{Owen:1998dk}
Owen B~J and Sathyaprakash B 1999 {\em Phys.Rev.\/} {\bf D60} 022002
  (\textit{Preprint} \eprint{gr-qc/9808076})

\bibitem{Babak:2006ty}
Babak S, Balasubramanian R, Churches D, Cokelaer T and Sathyaprakash B 2006
  {\em Class.Quant.Grav.\/} {\bf 23} 5477--5504 (\textit{Preprint}
  \eprint{gr-qc/0604037})

\bibitem{Harry:2009ea}
Harry I~W, Allen B and Sathyaprakash B~S 2009 {\em Phys. Rev.\/} {\bf D80}
  104014 (\textit{Preprint} \eprint{0908.2090})

\bibitem{Brown:2012qf}
Brown D~A, Harry I, Lundgren A and Nitz A~H 2012 {\em Phys. Rev.\/} {\bf
  D86}(8) 084017

\bibitem{Ajith:2012mn}
Ajith P, Fotopoulos N, Privitera S, Neunzert A and Weinstein A~J 2014 {\em
  Phys. Rev.\/} {\bf D89} 084041 (\textit{Preprint} \eprint{1210.6666})

\bibitem{Keppel:2013yia}
Keppel D 2013 {\em Phys.Rev.\/} {\bf D87} 124003 (\textit{Preprint}
  \eprint{1303.2005})

\bibitem{Keppel:2013uma}
Keppel D 2013  (\textit{Preprint} \eprint{1307.4158})

\bibitem{Privitera:2013xza}
Privitera S, Mohapatra S~R~P, Ajith P, Cannon K, Fotopoulos N, Frei M~A, Hanna
  C, Weinstein A~J and Whelan J~T 2014 {\em Phys. Rev.\/} {\bf D89} 024003
  (\textit{Preprint} \eprint{1310.5633})

\bibitem{Capano:2016dsf}
Capano C, Harry I, Privitera S and Buonanno A 2016  (\textit{Preprint}
  \eprint{1602.03509})

\bibitem{Buonanno:2009zt}
Buonanno A, Iyer B, Ochsner E, Pan Y and Sathyaprakash B 2009 {\em Phys.Rev.\/}
  {\bf D80} 084043 (\textit{Preprint} \eprint{0907.0700})

\bibitem{Brown:2012nn}
Brown D~A, Kumar P and Nitz A~H 2013 {\em Phys.Rev.\/} {\bf D87} 082004
  (\textit{Preprint} \eprint{1211.6184})

\bibitem{Nitz:2013mxa}
Nitz A~H, Lundgren A, Brown D~A, Ochsner E, Keppel D {\em et~al.\/} 2013 {\em
  Phys.Rev.\/} {\bf D88} 124039 (\textit{Preprint} \eprint{1307.1757})

\bibitem{Kumar:2015tha}
Kumar P, Barkett K, Bhagwat S, Afshari N, Brown D~A {\em et~al.\/} 2015
  (\textit{Preprint} \eprint{1507.00103})

\bibitem{Buonanno:1998gg}
Buonanno A and Damour T 1999 {\em Phys.Rev.\/} {\bf D59} 084006
  (\textit{Preprint} \eprint{gr-qc/9811091})

\bibitem{Pan:2009wj}
Pan Y, Buonanno A, Buchman L~T, Chu T, Kidder L~E {\em et~al.\/} 2010 {\em
  Phys.Rev.\/} {\bf D81} 084041 (\textit{Preprint} \eprint{0912.3466})

\bibitem{Damour:2012ky}
Damour T, Nagar A and Bernuzzi S 2013 {\em Phys.Rev.\/} {\bf D87} 084035
  (\textit{Preprint} \eprint{1212.4357})

\bibitem{Damour:2014sva}
Damour T and Nagar A 2014 {\em Phys.Rev.\/} {\bf D90} 044018 (\textit{Preprint}
  \eprint{1406.6913})

\bibitem{Capano:2013raa}
Capano C, Pan Y and Buonanno A 2013  (\textit{Preprint} \eprint{1311.1286})

\bibitem{Lyons:1900zz}
Lyons L 2008 {\em Ann. Appl. Stat.\/} {\bf 2} 887--915

\bibitem{Harry:2013tca}
Harry I~W, Nitz A~H, Brown D~A, Lundgren A~P, Ochsner E and Keppel D 2014 {\em
  Phys. Rev.\/} {\bf D89} 024010 (\textit{Preprint} \eprint{1307.3562})

\bibitem{Kehl:2014}
Kehl Marcel S 2014 {\em {Comparison of Fixed and Regenerated Template Banks for
  Compact Binary Coalescence Searches with advanced LIGO}\/} Master's thesis
  University of Toronto

\bibitem{Canton:2013joa}
Canton T~D, Bhagwat S, Dhurandhar S and Lundgren A 2014 {\em
  Class.Quant.Grav.\/} {\bf 31} 015016 (\textit{Preprint} \eprint{1304.0008})

\bibitem{Chatterji:2004qg}
Chatterji S, Blackburn L, Martin G and Katsavounidis E 2004 {\em Class. Quant.
  Grav.\/} {\bf 21} S1809--S1818 (\textit{Preprint} \eprint{gr-qc/0412119})

\bibitem{Robinet:2015om}
Robinet F 2015 {\em https://tds.ego-gw.it/ql/?c=10651\/}
  \urlprefix\url{https://tds.ego-gw.it/ql/?c=10651}

\bibitem{Brown:2004vh}
Brown D~A 2004 {\em {Searching for gravitational radiation from binary black
  hole MACHOs in the galactic halo}\/} Ph.D. thesis Wisconsin U., Milwaukee
  (\textit{Preprint} \eprint{0705.1514})
  \urlprefix\url{https://inspirehep.net/record/673638/files/arXiv:0705.1514.pdf}

\bibitem{FFTW05}
Frigo M and Johnson S~G 2005 {\em Proceedings of the IEEE\/} {\bf 93} 216--231
  special issue on ``Program Generation, Optimization, and Platform
  Adaptation''

\bibitem{Allen:1999yt}
Allen B {\em et~al.\/} 1999 {\em Phys. Rev. Lett.\/} {\bf 83} 1498
  (\textit{Preprint} \eprint{gr-qc/9903108})

\bibitem{Robinson:2008un}
Robinson C, Sathyaprakash B and Sengupta A~S 2008 {\em Phys.Rev.\/} {\bf D78}
  062002 (\textit{Preprint} \eprint{0804.4816})

\bibitem{Canton:2014ena}
Dal~Canton T {\em et~al.\/} 2014 {\em Phys. Rev.\/} {\bf D90} 082004
  (\textit{Preprint} \eprint{1405.6731})

\bibitem{Capano:2016uif}
Capano C, Dent T, Hu Y~M, Hendry M, Messenger C and Veitch J 2016
  (\textit{Preprint} \eprint{1601.00130})

\bibitem{Purrer:2015tud}
P{\"{u}}rrer M 2016 {\em Phys. Rev. D\/} {\bf 93}(6) 064041 (\textit{Preprint}
  \eprint{1512.02248})
  \urlprefix\url{http://link.aps.org/doi/10.1103/PhysRevD.93.064041}

\bibitem{Harry:2016ijz}
Harry I, Privitera S, Bohé A and Buonanno A 2016 {\em Phys. Rev.\/} {\bf D94}
  024012 (\textit{Preprint} \eprint{1603.02444})

\bibitem{Macleod:2015jsa}
Macleod D, Harry I~W and Fairhurst S 2016 {\em Phys. Rev.\/} {\bf D93} 064004
  (\textit{Preprint} \eprint{1509.03426})

\bibitem{Harry:2010fr}
Harry I~W and Fairhurst S 2011 {\em Phys. Rev.\/} {\bf D83} 084002
  (\textit{Preprint} \eprint{1012.4939})

\bibitem{Williamson:2014wma}
Williamson A, Biwer C, Fairhurst S, Harry I, Macdonald E, Macleod D and Predoi
  V 2014 {\em Phys. Rev.\/} {\bf D90} 122004 (\textit{Preprint}
  \eprint{1410.6042})

\end{thebibliography}
\end{document}